\documentclass[sigconf, nonacm]{acmart}
\usepackage{xspace} 
\usepackage{amsmath} 
\usepackage{makecell}
\usepackage{natbib}
\usepackage{multirow}
\setcitestyle{numbers,sort&compress}
\usepackage{enumitem}

\usepackage{listings}
\usepackage{soul}

\usepackage{graphicx}
\usepackage{subfigure}
\usepackage{pifont}

\usepackage{booktabs}
\usepackage{cleveref}
\Crefname{figure}{Fig.}{Figs.}
\Crefname{table}{Table}{Tables}

\newcommand{\tool}{\textsc{XRFix}\xspace}

\definecolor{mylightblue}{RGB}{222, 235, 247}
\definecolor{mygray}{RGB}{237, 237, 237}

\definecolor{codegreen}{rgb}{0,0.6,0}
\definecolor{codegray}{rgb}{0.5,0.5,0.5}
\definecolor{codepurple}{rgb}{0.58,0,0.82}
\definecolor{backcolour}{rgb}{0.95,0.95,0.92}
\definecolor{alizarin}{rgb}{0.82, 0.1, 0.26}
\definecolor{azure(colorwheel)}{rgb}{0.0, 0.5, 1.0}
\definecolor{brandeisblue}{rgb}{0.0, 0.44, 1.0}
\definecolor{darktangerine}{rgb}{1.0, 0.66, 0.07}
\sethlcolor{codegreen}

\makeatletter
\DeclareRobustCommand\onedot{\futurelet\@let@token\@onedot}
\def\@onedot{\ifx\@let@token.\else.\null\fi\xspace}

\def\eg{\textit{e.g}\onedot} 
\def\ie{\textit{i.e}\onedot} 
 
\def\etc{\textit{etc}\onedot}

\def\etal{\textit{et al}\onedot}
\makeatother

\usepackage{array}
\usepackage{colortbl}
\usepackage{mathtools}
\UseRawInputEncoding

\usepackage{framed}

\newenvironment{graybox}
  {\vspace{-0.2cm}\begin{framed}}
  {\end{framed}\vspace{-0.2cm}}

\newcommand{\answer}[2]{%
  \begin{graybox}
    \textbf{Answer to RQ#1:} #2
  \end{graybox}
}

\usepackage{amsmath,amsfonts}
\usepackage{algorithmic}
\usepackage{textcomp}

\ifodd 0
\newcommand{\hn}[1]{\textcolor{blue}{{#1}}} 
\else
\newcommand{\hn}[1]{#1}
\fi

\AtBeginDocument{%
  }

\begin{document}

\author{Jingwen Wu}
\affiliation{%
  \institution{Hong Kong Baptist University}
  \city{Hong Kong}
  \country{China}}
\email{csjwwu@comp.hkbu.edu.hk}

\author{Hanyang Guo}
\affiliation{%
  \institution{Sun Yat-sen University}
  \city{Zhuhai}
  \country{China}}
\email{guohy36@mail2.sysu.edu.cn}

\author{Hong-Ning Dai}
\authornote{Corresponding author.}
\affiliation{%
  \institution{Hong Kong Baptist University}
  \city{Hong Kong}
  \country{China}}
\email{henrydai@comp.hkbu.edu.hk}

\author{Xiapu Luo}
\affiliation{%
  \institution{The Hong Kong Polytechnic
University}
  \city{Hong Kong}
  \country{China}}
\email{csxluo@comp.polyu.edu.hk}

\renewcommand{\shortauthors}{Wu et al.}

\title[XRFix: Exploring Performance Bug Repair of XR Apps with LLMs]{XRFix: Exploring Performance Bug Repair of Extended Reality Applications with Large Language Models}


\renewcommand{\shortauthors}{Wu et al.}

\begin{abstract}
As an emerging technology, Extended Reality (XR) provides end-users with an immersive experience of interacting with virtual and physical environments. Unlike traditional software, the execution of XR applications involves more computationally complex operations, such as 3D scene rendering, real-time animation, and process simulations. \hn{Inefficient coding practices} during the software development of XR applications may cause various performance bugs,  degrading user experience and even causing motion sickness. \hn{Thus, it is an urgent need to develop an automated program repair (APR) framework for fixing performance bugs in complex XR programs. However, it is non-trivial to achieve this goal due to several technical challenges: (1) a lack of a real-world XR codebase and bug dataset, (2) no accurate bug detection tool, and (3) no effective bug-fixing tool designed for XR performance bugs. To tackle these challenges, we present a novel large language model-based framework, namely \tool, to repair performance bugs for open-source XR programs.} 
To overcome the first challenge, we first construct a corpus of domain-specific performance bugs built with a codebase from 23 open-source XR projects and a dataset of XR-related bugs containing 104 real-world bugs. 
To address the second challenge, we tailor two static analysis tools for accurately detecting bugs in both C\# scripts and asset files. 
For the third challenge, we design different prompts to instruct large language models (LLMs) to fix XR bugs in three types of bug scenarios with different complexities, i.e., single-line level, function level, and class level. 
We conduct extensive experiments on five off-the-shelf LLMs to evaluate the bug-fixing performance of \tool. \hn{We also compare our \tool with three state-of-the-art (SOTA) APR approaches.} Through static analysis, reference answer comparison, and manual inspection, \hn{we demonstrate that our \tool can effectively fix XR bugs, outperforming SOTA APR methods. For example, our \tool achieves 67.3\% fixing rate, i.e., 7.7\% higher than the second-best method.} 
We make \tool available on GitHub.\footnote{\url{https://github.com/wwwjwww/XRFix}}.

\end{abstract}

\maketitle

\vspace{-0.3cm}
\section{Introduction}
Recently, the prevalence of virtual reality (VR) and augmented reality (AR) devices has boosted the development of diverse VR and AR applications, consequently forming the extended reality (XR) ecosystem. 
XR applications provide users with an immersive experience of viewing and operating (even controlling) virtual objects and scenes through wearable XR headsets and peripheral devices. The recent advent of XR technologies has proliferated a broad spectrum of XR applications (apps) across diverse domains, such as smart manufacturing, entertainment, education, training, social media, and gaming~\cite{Trimananda2022,10.1145/3597503.3639082}. The number of XR users will reach more than 170 million, and the market size of XR hardware and software is expected to reach US\$ 3 trillion by the end of 2037~\cite{xrmarket}.

Unlike conventional mobile apps, XR apps typically consume much higher computation resources on GPU, CPU, and memory due to their 
extensive computation-complex operations, such as 3D scene rendering, real-time animations, and process simulations. Moreover, the limited computation resources of XR devices
can incur performance issues, such as animation lagging, video degradation, and frame discontinuity
~\cite{9402052,rzig2023virtual, 10.1145/3660803}. In contrast to conventional software, the performance issues of XR apps can severely affect user experience and even cause motion sickness or dizziness~\cite{10458372,tasnim2024investigating,ramaseri2025exploring}.   

Recently, lots of research endeavors have been made to address performance issues in XR apps. In particular, Nusrat \etal~\cite{9402052} presented an exploratory study on collecting open-source VR projects and investigating performance optimization commits. 
Bosco \etal~\cite{10174134} proposed a tool to detect typical bugs in VR apps. There are few studies on automatic program repair (APR) for XR apps, although APR is not a new topic in software engineering (SE). Many research efforts have been made to automate human-supervised program repair processes by employing advanced machine learning and deep learning techniques. These APR approaches include search-based APR~\cite{le2011genprog,jobstmann2005program}, template-based APR~\cite{le2016history, kim2013automatic, koyuncu2020fixminer, koyuncu2019ifixr}, neural machine translation (NMT)-based APR~\cite{tufano2019empirical,jiang2021cure,8827954}, \etc. Currently, various methods integrate LLMs' empirical knowledge into APR tasks~\cite{tufano2019empirical,xia2022less,peng2024domain,fan2023automated, Hammond2023}. Although these previous APR approaches demonstrate their effectiveness in conventional software, it is questionable to directly apply them to fix bugs in XR apps.
For example, 
XR performance bugs can be caused by inappropriate usage of computational resources 
during the entire life cycle. Moreover, 
\textit{inefficient~coding}~practices in script files and wrong settings of scenes in XR apps are also the root cause of bad simulations.
Therefore, it is an urgent need to develop an APR framework for fixing performance bugs in XR programs. 

\hn{However, it is \textit{non-trivial} to achieve this goal due to several technical challenges: (1) lacking a real-world XR codebase and bug dataset, (2) static tools dedicated to XR performance bug detection, and (3) an effective bug-fixing framework customized for XR programs. Motivated by recent advances in APR and large language models (LLMs), we present a novel LLM-based bug-fixing framework, namely \tool, to fix performance bugs in open-source XR programs.} To address the first challenge, we first collect 23 open-source XR projects developed by Unity (one of the most prevalent XR development frameworks with more than 60\% market share). We then summarize 10 types of XR-related bugs from related forums, MITRE's CWE database~\cite{cwe}, and related research studies. Next, we construct the XR codebase with real-world XR bugs. To tackle the second challenge, we develop a static bug detector for XR programs by defining new bug-detecting rules and integrating smell detectors into two customized static tools: CodeQL and UnityLint. To overcome the third challenge, we design different prompts for LLMs to deal with three types of bug scenarios, \ie, single-line level, function level, and class level. We conduct extensive experiments on five off-the-shelf LLMs, including both general-purpose LLMs (GPT-4 and GPT-3.5) and code LLMs (Code Llama, Deepseek-Coder, and StarChat-$\beta$) to evaluate the bug-fixing performance of \tool. Through static analysis, reference answer comparison, and manual inspection, we experimentally demonstrate that \tool outperforms SOTA APR methods, such as AlphaRepair, Fine-tuned CodeT5, and Self-Repair (GPT4o) in fixing XR performance bugs. 



The main contributions of this paper are summarized as follows.
\begin{itemize}[leftmargin=*,labelindent=0pt,noitemsep,topsep=0pt]
    \item To the best of our knowledge, this is the first work for performance bug fixing tasks in open-source XR programs based~on~LLMs. 
    \item We construct an XR open-source codebase and a real-world bug dataset. These real-world XR bugs spread all over the entire development life cycle of XR apps. 
    \item We integrate two customized static analyzers into \tool to effectively detect and localize Unity-related bugs across both C\# scripts and asset files by defining new bug-detecting rules.
    \item We design prompt templates to instruct state-of-the-art LLMs to fix bugs in three types of bug scenarios with different complexity levels: single-line level, function level, and class level.
    \item We conduct extensive experiments to evaluate LLMs in bug-fixing tasks through static analysis, reference answer comparison, and manual inspection. 
    Experimental results show LLMs' potential for fixing performance bugs in XR programs. \hn{Further, our \tool also outperforms existing SOTA approaches.}
\end{itemize}

\section{Background}

\subsection{XR Unity-related Bugs} \label{bug type}
\begin{figure}[t]
\centering
\setlength{\abovecaptionskip}{0.1cm}
\includegraphics[width=1\linewidth]{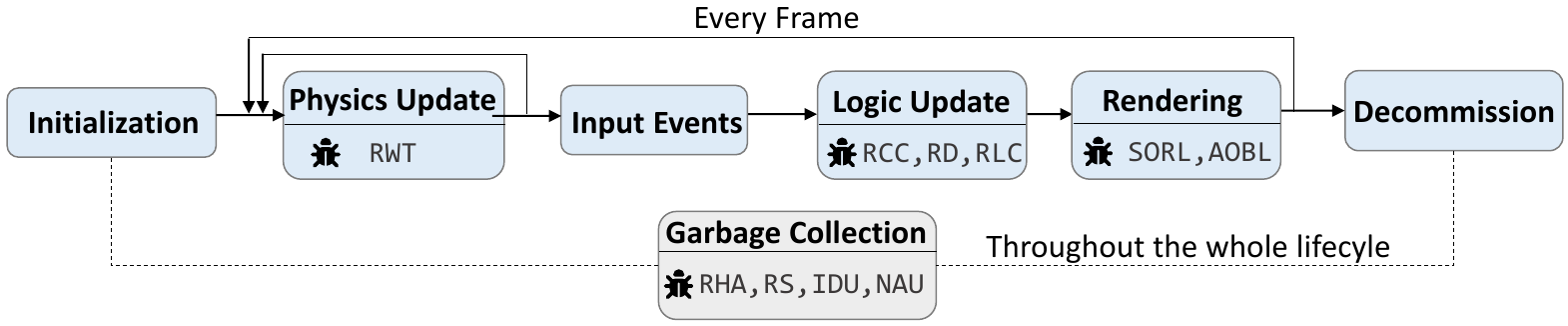}
\caption{
 XR-related bugs spreading all over the development Life-Cycle of Unity XR programs.}
\label{fig:Life_Cycle}
\centering
\vspace{-0.4cm}
\end{figure}

We mainly focus on XR apps built upon the Unity framework, which is one of the most dominant frameworks integrated with almost all existing XR platforms, such as ARKit, ARCore, Steam VR, HoloLens, \etc. In XR apps developed by the Unity framework, multiple scenes exist, where a \textit{scene} refers to the space where users interact. Game objects attached by C\# source code scripts defining their logic behaviors are the core components of these scenes. Life-cycle callback methods are invoked on each game object in the scene, potentially triggering diverse bugs. Figure~\ref{fig:Life_Cycle} shows the Unity development life cycle, which is composed of \textbf{Initialization}, \textbf{Physics Update}, \textbf{Input Events}, \textbf{Logic Update}, and \textbf{Rendering} processes. Specifically, Initialization is called before a scene is loaded or before updating the frame right after loading the game object. Thereafter, Physics Update, Input Events, Logic Update, and Rendering processes are sequentially invoked for each frame. \textbf{Decommission} is executed after the last frame of the scene or the object has been destroyed. During the entire lifecycle, \textbf{Garbage Collection} (GC) is a special mechanism, which can be triggered to free memory that Unity is no longer using. The GC process is quite unpredictable and costly. Inappropriate resource allocation may accelerate GC, affecting the animation due to heavy GC operations~\cite{9402052}.

\begin{table}[t]
\setlength{\abovecaptionskip}{0.1cm}
\caption{Bugs during Unity Development Life-cycle.}
\label{tab:info}
\centering

\scalebox{0.64}{
\setlength{\tabcolsep}{1mm}{
\begin{tabular}{lp{4.8cm}lp{4.9cm}}\toprule
\textbf{Stage} & 
\textbf{Bug\_Name} & 
\textbf{Abbr.}&
\textbf{Description}
\\
\midrule
\rowcolor{mylightblue}\multirow{2}{*}{\bf Physics Update} & \multirow{2}{*}{Rigidbody Weak Temporization} & \multirow{2}{*}{\texttt{RWT}} &  Change the position of Rigidbody object in \texttt{\small Update()} Function. \\
\midrule

\rowcolor{mylightblue}
 & \multirow{2}{*}{Redundant Condition Computation} & \multirow{2}{*}{\texttt{RCC}} &Remove the conditions that are constant. \\
 \rowcolor{mylightblue}& \multirow{2}{*}{Redundant Logic Checks} & \multirow{2}{*}{\texttt{RLC}} & Remove redundant non-short circuit operators. \\
 \rowcolor{mylightblue} \multirow{-5}{*}{\bf Logic Update} & Resource Deadlock & \texttt{RD} & Avoid using ``this'' in a ``lock'' statement \\
\midrule
\rowcolor{mylightblue}  & Static Object uses Real-time Lighting & \texttt{SORL}
& Excessive rendering \\
\rowcolor{mylightblue}\multirow{-2}{*}{\bf Render} 
 & Animation Object uses Baked Lighting & \texttt{AOBL} & Rendering Distortion 
\\\midrule
 \rowcolor{mygray} & \multirow{2}{*}{Redundant Heap Allocation} & \multirow{2}{*}{\texttt{RHA}} & Heap allocation of useless contents of a collection. \\
 \rowcolor{mygray}& Redundant Select &  \texttt{RS} & Redundant select method in LINQ. \\
 \rowcolor{mygray} & \multirow{2}{*}{Instantiate and Destroy in Update} & \multirow{2}{*}{\texttt{IDU}} & Instantiate and Destroy objects in \texttt{Update()} Function. \\
 \rowcolor{mygray} \multirow{-4}{*}{\bf GC}  & \multirow{2}{*}{New Allocator in Update} & \multirow{2}{*}{\texttt{NAU}} & Use \texttt{new()} for memory allocation in \texttt{Update()} function. \\
\bottomrule
\end{tabular}}}
\vspace{-0.4cm}
\end{table}

\begin{figure}[t]
\vspace{-0.2cm}
\setlength{\abovecaptionskip}{0.1cm}
\centering
\includegraphics[width=\linewidth]{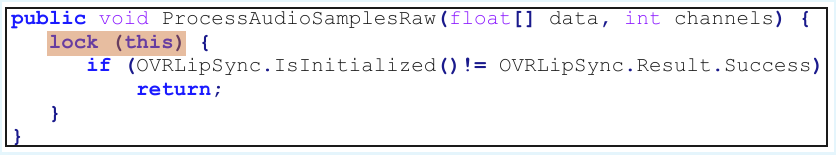}
\caption{
Example of RC. Using ``\texttt{\small this}'' in a ``\texttt{\small lock}'' statement could cause resource deadlock..}
\label{fig:code2}
\centering
\vspace{-0.4cm}
\end{figure}

We mainly consider ten bugs during the Unity development's life-cycle from VR-related forums, MITRE's Common Weakness Enumeration (CWE) database, and related studies. These bugs are distributed in Physics Update, Logic Update, and the Render process during the entire life-cycle. Table~\ref{tab:info} summarizes these bugs. We next briefly elaborate on these XR bugs in the life cycle.
\begin{itemize}[leftmargin=*] 
    \item [1)] \textbf{Physics Update} contains the following bug. \textit{Rigidbody Weak Temporization} (RWT) occurs when a Rigidbody object's settings are changed in the \texttt{\small Update()} function, leading to inaccurate physics simulations. In Unity official guidance, the \texttt{\small FixedUpdate()} function is recommended to apply forces to the rigidbody object and control it in a physically realistic way.

\item [2)] \textbf{Logic Update} contains the following bugs. (i) \textit{Redundant Condition Computation} (RCC) occurs when a condition always evaluates to true or always evaluates to false, thereby resulting in redundant computation.
(ii) \textit{Redundant Logic Checks} (RLC) occurs when non-short circuit operators are used, thereby degrading the efficiency of the program. An example of RLC is shown in Listing~\ref{fig:code1}. (iii) \textit{Resource Deadlock} (RD) occurs when  ``this'' reference is used in a ``lock'' statement, consequently leading to inefficiency or deadlock because of other classes attempting to lock the object. Figure~\ref{fig:code2} presents an example of RD.

\begin{figure}[t]
\vspace{-0.2cm}
\setlength{\abovecaptionskip}{0.1cm}
\centering
\includegraphics[width=\linewidth]{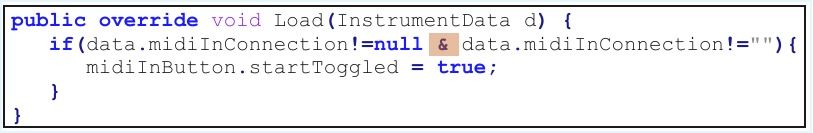}
\caption{
Example of RLC (``\&'' is a non-short circuit operator to cost heavier computation than  "\&\&").}
\label{fig:code1}
\centering
\vspace{-0.4cm}
\end{figure}



\item [3)] \textbf{Render} contains two bugs. (i) \textit{Static Object uses Real-time Lighting} (SORL) occurs when a static object emits a real-time light, resulting in excessive rendering. (ii) \textit{Animation Object uses Baked Lighting} (AOBL) occurs when an object associated with the animation script emits a baked light, leading to distorted rendering.

\end{itemize}

Besides the above bugs in Unity's development life cycle, there are several bugs in the GC process. (i) \textit{Redundant Heap Allocation} (RHA) takes place when the contents of a collection are never used, incurring additional performance overhead. (ii) \textit{Redundant Select} (RS) occurs when passing an identify function to LINQ's \texttt{\small Select()} method yields the same sequence, on which ``\texttt{\small Select()}'' is redundantly called. (iii) \textit{Instantiate and Destroy Objects in Update} (IDU) occurs when game objects in \texttt{\small Update()} function are instantiated and destroyed, 
resulting in frequent heap allocation because \texttt{\small Update()} is called each frame. (iv) \textit{New Heap Allocation in Update} (NAU) occurs when \texttt{\small new()} is used to invoke heap allocation in the \texttt{\small Update()} function. Specifically, both two bugs from the Render process are located in Unity's asset files\footnote{Asset files regulate the graphics behaviors in Unity (\eg, \texttt{\scriptsize.3ds}
and \texttt{\scriptsize.fbx} files for 3D models, \texttt{\scriptsize.shader} files for lighting
effects, \texttt{\scriptsize.mat} files for material textures, and \texttt{\scriptsize.unity} files
for scene settings).}, while other bugs exist in script files\footnote{Script files (typically in
C\#) that are source code files in XR apps are attached to GameObjects to define their logic behaviors.}.

\vspace{-0.3cm}
\subsection{Static Code Analysis}
Since static code analysis can analyze source codes without executing them, we also consider several static analysis tools in our framework.
To efficiently extract functions in uncompiled scripts, we adopt Tree-sitter~\cite{tree-sitter} during both data collection and the patch-merging procedure. 
Tree-sitter generates parsers to create syntax trees for specific programming languages. In \tool, we use Tree-sitter C\# grammar to generate the C\# parser\footnote{\url{https://github.com/tree-sitter/tree-sitter-c-sharp}}. Then, we extract functions using the AST from Tree-sitter and design rules based on bug patterns to identify different structures of bug scenarios. Tree-sitter helps us collect various bug types and integrate LLM-generated fixes into the original repository.

Inspired by recent advanced static code analysis tools in identifying bugs, we customize two SOTA static code analysis tools: UnityLint~\cite{10174134} and CodeQL~\cite{codeql} to assist in Unity-related bug detection.
UnityLint adopts the Roslyn compiler~\cite{roslyn} and its API to analyze source codes and asset files. It defines detection rules for bad smells by querying JSON files to search for specific variables or methods.
Despite the predefined 18 bad Unity smells, UnityLint has its limitations in detecting complex data flows. Moreover, UnityLint is not specially designed to construct call graphs precisely. 
In contrast, CodeQL can statically analyze source code in script files with the support of complex data flows. 
CodeQL lets users retrieve codes by writing queries using declarative Query Language (QL) to find different vulnerabilities. For C\# source codes, CodeQL requires Compile \& Build tasks to acquire the program's derived information, such as type hierarchy or macro expansions. To define the Unity-related bugs, we adopt CodeQL's global data flow analysis. By defining sinks and sources in queries, CodeQL investigates data flows between functions and object properties across the whole project. Although CodeQL is efficient at detecting complex data flows, it only supports C\# source codes in XR Unity repositories. 

Considering the pros and cons of UnityLint and CodeQL, we integrate them in our \tool together to detect Unity-related bugs.
\vspace{-0.2cm}
\subsection{Automated Program Repair}
APR tools can generate plausible patches for error codes, given their fault locations. Conventional APR tools typically exploit search-based~\cite{ke2015repairing,xin2017leveraging}, constraint-based~\cite{xuan2016nopol, nguyen2013semfix}  and template-based~\cite{le2016history, kim2013automatic, koyuncu2020fixminer, koyuncu2019ifixr} techniques. To tackle traditional APRs' limitations, 
learning-based APR techniques have emerged. Among them, NMT-based and LLM-based tools are the most prevalent ones~\cite{sutskever2014sequence}. Although NMT-based APR tools can fix more bugs than traditional APR, they are still limited in terms of types and the number of bugs.

\begin{table}[t]
\setlength{\abovecaptionskip}{0.1cm}
\caption{LLMs adopted in this paper}
\label{tab:llm}
\centering
\scalebox{0.7}{
\begin{tabular}{llp{2cm}lp{2.8cm}}
\toprule
\textbf{Model}  & \multicolumn{1}{c}{\textbf{\#Params}} & \multicolumn{1}{c}{\textbf{Context Window}} & \textbf{Type} & \textbf{Cut-off Dates} \\ \midrule
Code Llama & 7B & 16K & Code LLMs&2022.09\\
Deepseek-Coder & 6.7B & 16K & Code LLMs&2023.03\\
StarChat-$\beta$ & 16B & 8K & Code LLMs&N.A. (Before 2023.05)\\
GPT-3.5-Turbo & 175B& 16K & General LLMs&2021.09\\  
GPT-4o & 1800B & 128K & General LLMs&2023.10\\ 
\bottomrule
\end{tabular}}
\vspace{-0.5cm}
\end{table}

LLMs, pre-trained on a large amount of text data, have achieved impressive performance on diverse language-related and code-related tasks. Those code-related tasks include code generation~\cite{gu2023llm}, code summarization~\cite{ahmed2024automatic}, code repair~\cite{kulsum2024case, parasaram2025fact}, test case generation~\cite{schafer2023empirical}, and so on. Recently, prompt engineering has emerged as a method to guide LLMs in executing tasks by clearly specifying task descriptions and expected results, eliminating the need for fine-tuning. Since LLMs are trained on a massive amount of open-source codes, it is promising to apply LLMs to APR. 

There are three common methods to integrate LLMs with APR: fine-tuning~\cite{mashhadi2021applying,huang2025comprehensive}, few-shot learning~\cite{fan2023automated,nashid2023retrieval}, and zero-shot learning~\cite{xia2022less,peng2024domain,Hammond2023}. 
Our goal is to investigate the prompt engineering potential of various LLMs in fixing XR bugs with zero-shot and one-shot learning. Specifically, we 
design different types of prompts to instruct LLMs with role-based prompts.
Table~\ref{tab:llm} summarizes five SOTA LLMs adopted in this paper. Code Llama, Deepseek-Coder, and StarChat-$\beta$  are Code LLMs, which were specially trained to conduct code-related tasks, while GPT-3.5 and GPT-4o are general LLMs, which are popular in diverse domains. The \textit{context window} of LLMs specifies the number of tokens that the model can take as input when generating responses. We must determine the maximum number of tokens that can be generated per instruction (denoted by ``max\_token'') in our experiments (to be given in \S~\ref{sec:exp}). \hn{ We have also compared LLMs' repairing abilities with three SOTA APR methods in \S~\ref{compare}}.

\vspace{-0.3cm}
\section{Methodology}
\begin{figure*}[t]
\centering
\setlength{\abovecaptionskip}{0.1cm}
\includegraphics[scale=0.49]{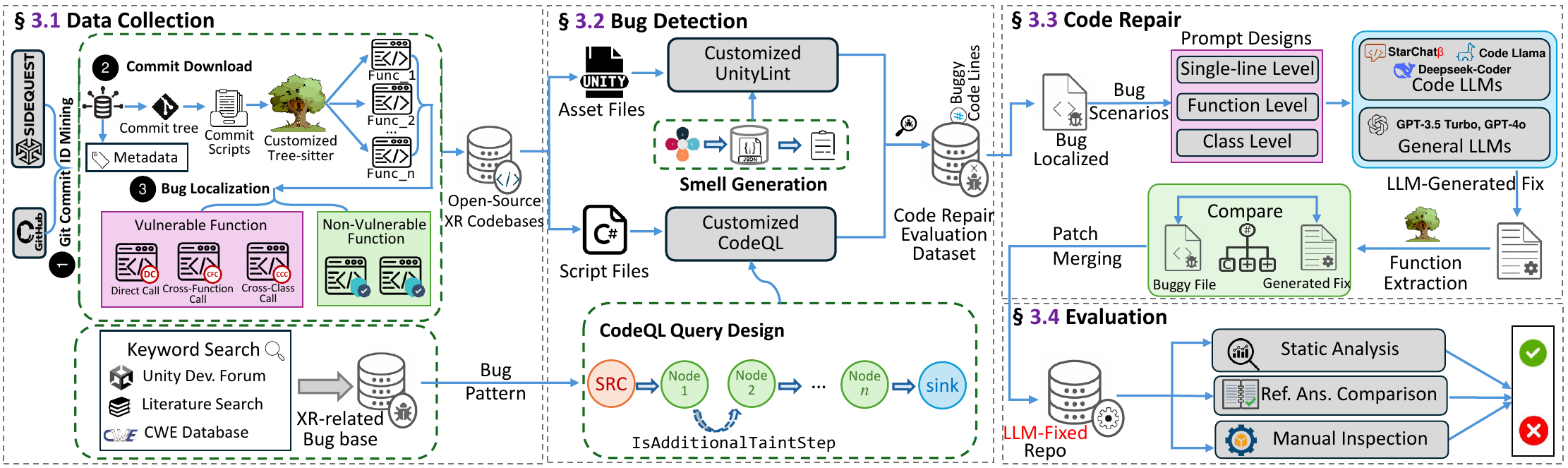}
\caption{
\hn{Workflow of \tool.}}
\label{fig:Code_Repair_Framework}
\centering
\vspace{-0.2cm}
\end{figure*}

Figure~\ref{fig:Code_Repair_Framework} depicts the workflow of the proposed \tool. First, we collect open-source XR repositories to construct the evaluation dataset. Second, we customized static analysis tools by adding newly designed rules. Third, after obtaining detailed information about these bugs, we design different types of prompts for state-of-the-art LLMs to fix XR bugs. Last, after constructing ``LLM-Fixed'' repositories by incorporating fixed codes generated by LLMs, we conduct static analysis, reference answer comparison, and manual inspection to verify whether a bug is successfully fixed or not. 

\vspace{-0.3cm}
\subsection{Data Collection}
We collect open-source XR projects from SideQuest (i.e., one of the largest XR app markets) and GitHub. For Unity-developed apps from SideQuest, we further gain 25 popular open-source XR projects from diverse repositories (including GitHub, GitLab, and other open-source repository platforms), each with a minimum of 2,300 downloads. We call these open-source XR projects \textit{SideQuest projects}.
For XR-related projects from GitHub, we use GitHub APIs to identify relevant commit IDs through keyword searching. Initially, to limit the repository size, we focus on XR repositories by using keywords, such as ``Augmented Reality'', ``Virtual Reality'', ``Mixed Reality'', ``Extended Reality'' and their acronyms (``AR'', ``VR'', ``MR'', ``XR''). We further check these repositories to ensure they were developed by Unity. In this way, we roughly select 72 XR-related repositories, which are called \textit{GitHub projects}. Adding the 25 projects, we obtain 97 XR repositories.


Next, we collect XR-related bugs in real-world software development. We first summarize XR-related keywords including ``AR'', ``VR'', ``Oculus'', ``Unity'', 
``Oculus Quest'', 
and performance-related keywords such as ``compatibility'', ``performance'', ``speed up'', ``accelerate'', 
``latency'', ``optimize'', ``efficient'', ``frame rate'', 
``render'', 
\etc. Second, we search XR-related online forums, related research papers, and MITRE's CWE database by using these keywords. During this process, two experienced XR researchers are invited to review all the results and finally filter out 33 bug candidates. To better fit into the APR task, we analyze each bug pattern and consider whether it is decisive to fix it by LLMs without human intervention. 
Eventually, we identify 10 typical XR bugs as defined in \S~\ref{bug type}.

\newcommand{\circledNum}[1]{\ooalign{
  \hfil\raisebox{0.1em}{\textcircled{\small #1}}\hfil\cr
}}

\noindent\circledNum{1}{\textbf{ Git Commit ID Mining}}. To collect bugs from real-world scenarios, we first apply keywords specific to different bugs to examine the commit changes in all the above 97 open-source projects. For instance, for \texttt{\small IDU}, we choose commit changes that contain the keywords ``Instantiate'', ``Destroy'', and ``Update''. During this process, we exclude 35 repositories that lacked relevant commit contents.

\noindent\circledNum{2}{\textbf{ Commit Download}}. After acquiring the potential commit IDs, we download the metadata and the corresponding GitHub repositories, switching to specific commit branches. 

\noindent\circledNum{3}{\textbf{ Bug Localization}}. We adopt UnityLint and CodeQL for bug localization, where UnityLint is used to locate bugs in asset files. For bugs in C\# script files, we adopt a curated data collection approach by using Tree-sitter and CodeQL together.

First, we use Tree-sitter to separate functions by constructing a concrete syntax tree for a source file. Then, we develop customized syntax rules to detect buggy code variations that are specific to XR characteristics.. 
 We categorize function call complexities into three levels according to XR's characteristics: (1) direct call (DC), (2) cross-function call (CFC), and (3) cross-class call (CCC), as illustrated in Figure~\ref{fig:functioncall}. In the DC scenario, \texttt{\small Instantiate()} is invoked directly within \texttt{\small Update()}. In the CFC scenario, while \texttt{\small Instantiate()} is invoked by \texttt{\small CreateObject()}, \texttt{\small CreateObject()} is called inside the \texttt{\small Update()} function. In the CCC scenario, a \texttt{\small GetComponent()} method is used to create a reference to another component of type \texttt{\small CheckLight}, which then invokes the \texttt{\small CreateObject()} function from the \texttt{\small CheckLight} component containing the \texttt{\small Instantiate()} function. Tree-sitter identifies these scenarios using our designed syntax rules, narrowing them down to 30 repositories for further analysis.

Since CodeQL can derive complex dataflows depending on the compilation process, we utilize it to validate these bugs from target repositories. There are three steps to perform CodeQL analysis for C\# codes of Unity projects. First, we construct the CodeQL database by compiling C\# projects. 
Second, we run CodeQL queries against the database. Last, we adopt CodeQL to interpret query results. Notably, we create CodeQL queries to analyze data flows and identify each bug across various code scenarios (details given in \S~\ref{bugdetect}).\hn{ We determine each bug scenario based on the locations of the source and sink as identified by CodeQL. For example, in the case of cross-class call, the source and sink must be located in two different classes.} After excluding repositories that failed to compile, we eventually obtain 23 open-source XR projects: 10 SideQuest projects and 13 GitHub projects.
Table~\ref{tab:app} shows the descriptive statistics for all open-source XR projects. 

In summary, we construct a codebase consisting of 23 open-source XR repositories.
During data collection, Tree-sitter is utilized to streamline the selection of target repositories. CodeQL queries are utilized to extract comprehensive data flow information. Additionally, UnityLint is adopted to detect bugs in asset files.

\begin{table}[t]
\vspace{-0.3cm}
\setlength{\abovecaptionskip}{0.1cm}
\caption{Descriptive statistics for Open-source XR Projects}
\label{tab:app}
\centering
\scalebox{0.7}{
\begin{tabular}{lllll}
\toprule
\textbf{Project\_IDs} & \textbf{\# of Repo}  & \textbf{Date Range} & \textbf{Source} & \textbf{Avg. LoC(C\#)}\\ \midrule
a-w & 23 & 2017-2024 & GitHub, GitLab, and others & 90.1k\\
\bottomrule
\end{tabular}}
\vspace{-0.5cm}
\end{table}

\vspace{-0.3cm}
\subsection{Bug Detection}
\label{bugdetect}

We adopt CodeQL and UnityLint to detect XR-related bugs. First, we utilize existing queries predefined by CodeQL to detect bugs, such as \texttt{\small RCC}, \texttt{\small RLC}, \texttt{\small RD}, \texttt{\small RHA}, and \texttt{\small RS}. Meanwhile, we tailor UnityLint to detect \texttt{\small SORL} and \texttt{\small AOBL} bugs by extracting information from Unity asset files. Notably, for bug \texttt{\small NAU}, \texttt{\small RWT} and \texttt{\small IDU} (not existing in CodeQL queries), we customize CodeQL by designing new types of queries. We next describe the details as follows.

\begin{figure}[t]
\vspace{-0.3cm}
\setlength{\abovecaptionskip}{0.1cm}
\centering
\includegraphics[width=\linewidth]{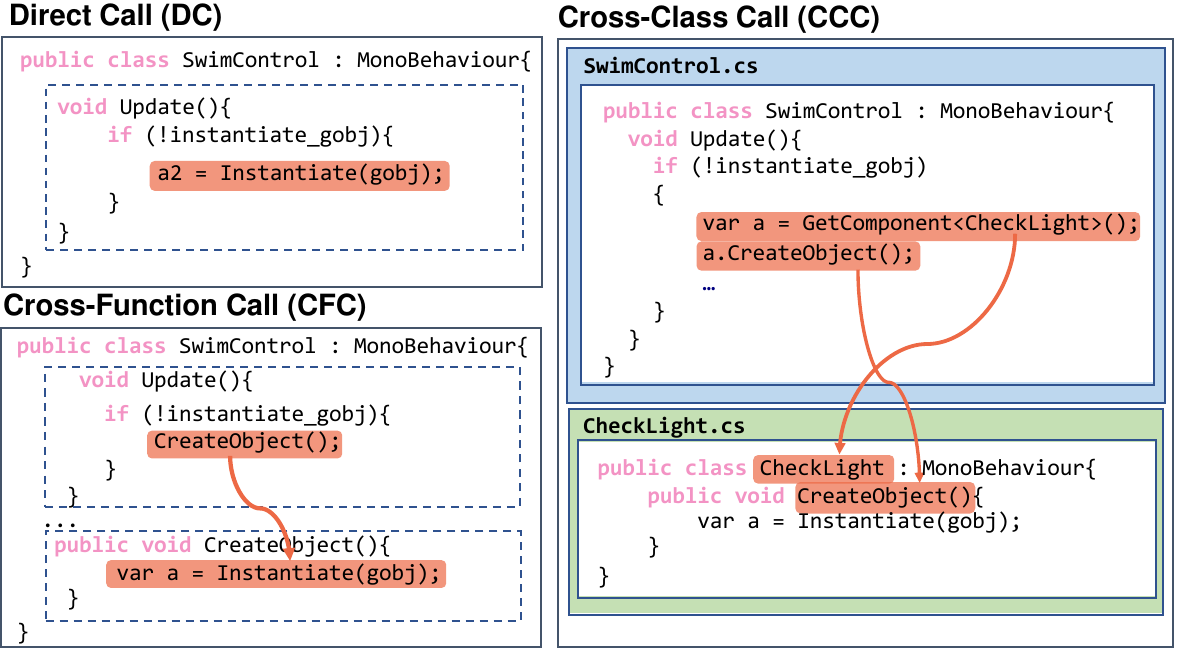}
\caption{
Three different function calls for bug \texttt{\small IDU}.}
\label{fig:functioncall}
\centering
\vspace{-0.5cm}
\end{figure}

\textbf{Smell Detection of UnityLint.} We tailor UnityLint to detect smells in Unity asset files. First, we utilize the \textit{Unity Data Analyzer} module to extract the information in asset files, which outputs a JSON file containing all the information. Second, we use the \textit{Meta Smell Analyzer} module to read the JSON file and apply \textit{existing} detection rules to locate \texttt{\small SORL} and \texttt{\small AOBL} in asset files. We combine these two components and recompile the repository based on its source code to create a new release as the customized UnityLint.

\begin{figure}[t]
\vspace{-0.2cm}
\setlength{\abovecaptionskip}{0.1cm}
\centering
\includegraphics[width=1.05\linewidth]{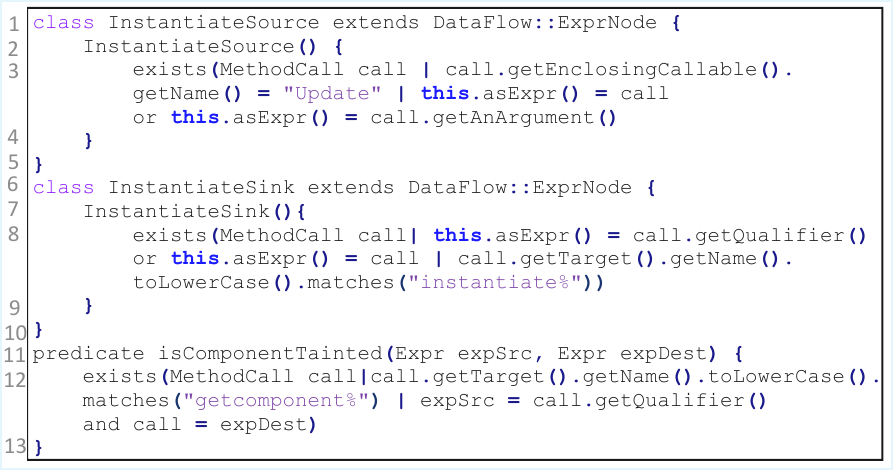}
\caption{
Query detecting if there exist Instantiate/Destroy function calls in the Update() function.}
\label{fig:code3}
\centering
\vspace{-0.4cm}
\end{figure}

\textbf{CodeQL Query Design.} 
We mainly use \textit{existing} bug detection rules predefined in CodeQL to locate those XR bugs predefined in CWE databases. For other bugs that require new rules to detect specific data flows, we define new queries to track these data flows or functions, as \textit{customized} rules. In particular, we create detection rules for \texttt{\small RWT}, \texttt{\small IDU}, and \texttt{\small NAU} in CodeQL. Take \texttt{\small IDU} as an example (Figure!\ref{fig:code3}). We define all method calls inside Unity's inherent functions being called each period, such as \texttt{\small Update()} function, as the predicate of sources (lines~1-5), and method calls that match ``\texttt{\small \textcolor{codepurple}{instantiate}}'' as the sink (lines 7-10). 
Since CodeQL does not support identification of Unity's special function to reference from another component, we define an additional path (lines 11-13) from the source to method calls matching ``\texttt{\small \textcolor{codepurple}{getcomponent\%}}''. Similarly, we can detect the \texttt{\small Destroy()} function inside the \texttt{\small Update()} function in this way.

After applying static analysis tools to all our collected repositories, we scrutinize the bug scenarios and curate our code repair evaluation dataset. 
In summary, we curate 104 bugs with three different scenarios. The dataset includes detailed bug information, such as each XR bug's location, name, and description. To evaluate the program repairing capability of LLMs, each bug is associated with one or more reference answers as ground truth answers. Reference answers are given by two authors with a professional background in Unity development (details given in \S~\ref{subsec:rq3}). Initially, they independently write reference answers based on official Unity guidance, and then collaborate to discuss and refine their answers.

\vspace{-0.3cm}
\subsection{Program Repair}
\label{sec:programrepair}
After constructing the XR code dataset for program repair, we extract code lines concerning different bug scenarios. We then design four prompt templates to instruct state-of-the-art LLMs for generating code fixes. Next, we replace buggy codes with patch code fixes for further evaluation. We elaborate on bug scenarios and prompt designs as follows.

\textbf{Bug Scenarios.} We categorize XR bugs into three types of APR scenarios: (i) \textit{single-line level} bug containing a single error line within a function can be fixed by generating (replacing/adding) a single line, (ii) \textit{function-level} bug containing consecutive code lines in one or multiple functions can be fixed by generating (replacing/adding) single or multiple functions within the same class, and (iii) \textit{class-level} bug containing functions from another class can be fixed by generating new functions. Figure~\ref{fig:prompt_head} depicts an example of our designed prompt for class-level bugs.

\begin{figure}[t]
\vspace{-0.3cm}
\setlength{\abovecaptionskip}{0.1cm}
    \centering
    

    
    
    \scalebox{0.7}{
    \begin{tabular}{p{10cm}}
    \hline
    You're an Automated Program Repair tool.  The following C\# code is based on Unity Development. Your task is to fix the code under the 'FIXED CODE:' area. Please only change the code from [related script].
    \\
    // code in [\textbf{error script}]:
    
    \color{codegreen}[prefix]
    \\
    // \color{alizarin}[buggy caller function]
    \\ 
    \\
    // Here's the definition of a function call in another class.
    \\
    //  code in [\textbf{related script}]
    \\
    // \color{brandeisblue}[callee function]
    \\
    // FIXED CODE:
    \\
    \hline
    \end{tabular}    \label{fig:single_prompt}}
\caption{Prompt for class-level bug using "//" in script files.}
\label{fig:prompt_head}
\centering
\vspace{-0.3cm}
\end{figure}

\textbf{Prompt Design.} We design different prompt templates to fix the above types of bugs. For a single-line level bug, we include the code snippet from the prefix to the buggy line. For a function-level bug, we include the entire buggy function as well as other contents related to the buggy function in the prompt. 
As for the class-level bug, we include all related functions in the prompt. Specifically, the related functions include the entire buggy function (\ie, caller) from the source class and its callee function from another class.

We then examine diverse instruction variants based on the above three templates to guide bug~repairing. For guidance, we consider recommendations from LLM documentations, e.g.,~\cite{guidance}. These templates differ from different prompt components with a varied amount of contexts:  basic instructions,  \textit{bug instructions},  \textit{fix instructions}, \textit{code examples} of bugs and fixes based on prior studies of program repair. First, we use ``\textit{You're an Automated Program Repair tool.}'' as the default system message. Then, we indicate that the code contains a bug and ask the model to provide a fix. For C\# script files, \textit{bug instruction} contains ``BUG:'' and ``MESSAGE:'' to include the bug's name and descriptions. We next include error code lines in a commented-out style, which is prevalent in code commits of open-source XR bug-fixing processes. Thereafter, we specify ``FIXED CODE:'' as the \textit{fix instruction}. \hn{As for \textit{code examples}, we collect the specific types of bugs and fixes from GitHub and Unity forums as examples to instruct LLMs on the expected output, suggesting each bug's common patterns. We include this section after the task description with a separation from the following vulnerable code~lines.} The fix instruction is preceded by suggestions to assist in fixing. Moreover, we also extend prompt templates from C\# script files to asset files. 
Specifically, we use plain text to include buggy contents rather than a commented-out style due to YAML's language characteristics. Thus, there is no prompt d for asset files. Table~\ref{tab:prompt} summarizes~prompt~templates for XR program repair. We further explore each prompt component's effects in \S~\ref{abla}.

\vspace{-0.3cm}
\subsection{Evaluation}
After obtaining the generated response from LLMs, we replace the buggy code with the repaired code in their source code. We then evaluate the correctness and reliability of bug fixes. In particular, we introduce three methods for evaluation: (1) static analysis, (2) reference answer comparison, and (3) manual inspection. 

With regard to static analysis, we adopt the fixed rate and percentage of plausible patches to evaluate the results. 
Furthermore, we conduct reference answer comparisons to evaluate the effectiveness of bug fixes. In particular, one author with professional background provides one or more reference answers for each bug scenario. For bugs in C\# script files, we then adopt \textit{CodeBLEU} metric to compare the similarity between reference answers and patches from LLMs.
For bugs in asset files, researchers have scored ``0'' or ``1'' to compare the LLM-generated answers with reference answers, where ``1'' represents the same answers. We utilize the average scores of responses as their similarity scores.

Finally, we conduct a manual inspection by randomly selecting a large number of plausible responses to examine the correctness of bug fixes. Thereafter, we select the bug fixes generated by LLMs (\eg, GPT-4o) and replace the original program (either C\# scripts or asset files) with fixes. We compile and execute the entire project to test its full-fledged function and reliability.
\begin{table}[t]
\vspace{-0.1cm}
\setlength{\abovecaptionskip}{0.1cm}
\caption{Template Variants for XR Program Repair}
\label{tab:prompt}
\centering
\scalebox{0.7}{
\setlength{\tabcolsep}{1mm}{
\begin{tabular}{cp{5.25cm}p{5.25cm}}
\toprule
\textbf{TPL\_ID}  & \textbf{Templates for Script Files} & \textbf{Templates for Asset Files} \\ \midrule
a & Basic Prompt - After instructing the LLMs on the detailed task description, the prompt will contain the `commented-out' version (using //) of vulnerable code/function body followed by a "FIXED CODE:". 
&  Basic Prompt - After instructing the LLMs on the detailed task description, the prompt will contain the error settings from asset files followed by a "FIXED:". \\ \midrule
b & Basic Prompt + \textbf{Bug Instruction} - Same as a, but add a comment `BUG: [error name]' and `MESSAGE: [error message]' to include bug instruction.
& Basic Prompt + \textbf{Bug Instruction} - Same as a, but add the bug's name and description in front of the error settings.\\ \midrule
c & Basic Prompt + Bug Instruction + \textbf{Fix Instruction} - Same as b, but adds a comment of fix instruction before the line of "FIXED CODE:". 
& Basic Prompt + Bug Instruction + \textbf{Fix Instruction} - Same as b, add the suggestions to fix the error in front of the error settings.\\ \midrule
d & Basic Prompt + Bug Instruction + Fix Instruction with \textbf{Alternative Comment Style} - Same as c, but commented in alternative comment style using /* and */ rather than //. 
& N.A. \\ \midrule
e & \hn{Basic Prompt + Bug Instruction + \textbf{Code Examples} of Specific Bug and Fix  - Same as b, add code examples of bug and fix of target bug type.} 
& \hn{Basic Prompt + Bug Instruction + \textbf{Code Examples} of Specific Bug and Fix - Same as b, add examples of bug and fix before the target buggy code.}\\
\bottomrule
\end{tabular}}}
\vspace{-0.53cm}
\end{table}

\vspace{-0.3cm}
\section{Experiment}\label{sec:exp}

\subsection{Implementation and Experiment Details} 
\textbf{Experiment Settings.} We implement \tool by Python. Patches are generated on a server with Ubuntu 20.04.6 LTS and equipped with two NVIDIA Ampere A100 GPUs, each with 80 GB of memory. We conducted our manual inspection procedure on the Unity engine on an Intel Core i7-4770K CPU @ 3.50GHz. 
As a base case, we extract the original buggy code from the constructed bug repair dataset and derive a prompt to obtain fixes by querying different LLMs. During the bug fix procedure, we leverage LLMs' advantages by different methods. For GPT-3.5-Turbo and GPT-4o, we adopt the official API endpoint provided by ChatGPT to generate code patches. For Code Llama, StarChat-$\beta$, and Deepseek-Coder, we use HuggingFace~\cite{huggingface} to load model weights and generate outputs. For each chosen prompt, we manually examine a few alternative approaches with selected bugs via the OpenAI Playground after following the best-practice guide~\cite{playground}. We choose a sampling temperature of top$\_p$ = 1 to get a diverse set of potential fixes aligning with the default settings across most LLMs. We query an LLM five times for each bug scenario. 

\textbf{Code Reduction and Patch Merging.} Notably, our prompt might be too large to present in its entirety due to the context window limitations of each model (as shown in Table~\ref{tab:llm}) due to the characteristics of real-world bugs.
To address this issue, we begin our prompt with a list of C\# \texttt{\small using} directives. Then, we skip top-level statements to the beginning of the buggy function. Next, we include code lines according to our different prompt templates. For a fair comparison, we let the $\max\_\operatorname{tokens}$ of all LLMs be 4,000 due to context window limit. 

Different bug scenarios may cause LLMs to generate new code lines or functions. \hn{To merge them with the original code, we use Tree-sitter to extract function definitions and key variable declarations from each response and compare them with the original one, as shown in Figure~\ref{fig:Code_Repair_Framework}. We compare the buggy file's extracted AST features with the generated fix. If the response updates any bug-free functions, we replace the originals with the generated ones. We also replace buggy functions with their LLM-generated versions. However, if the LLM only provides code lines instead of full functions, we insert these lines right after the commented buggy function. After making these changes, we add the remaining code. This process produces a potentially LLM-fixed repository (called a \textit{plausible} fix).} We carefully combine all content to ensure reliable evaluation results.

\textbf{Experiment Evaluation.} To evaluate our experiment, we first use static analyzer tools to identify whether the bug has been fixed. 
For bugs in C\# script files, it is convenient for us to determine whether the answer code is compilable by checking the existence of the CodeQL database.  
Next, we compare the similarity between the generated code lines and our reference answer for all types of bugs. If more than one reference answer is provided, we always choose the one with the highest scores by comparing these responses with reference answers. Further, we manually assess these responses by choosing the bug-fixed projects to build and run them on the Unity platform. Each bug is considered to be fixed if any one of the samples has passed the evaluation tests.

\vspace{-0.4cm}
\subsection{Evaluation Metrics}\label{metrics}
To evaluate \tool, we mainly consider two sets of evaluation metrics: 1) For evaluating the effectiveness of bug detection, we adopt \textbf{precision} to quantify the proportion of warnings reported by our customized static analysis tools that precisely identify bugs.
2) For evaluating the effectiveness of program repair, we have:
\begin{itemize}[leftmargin=*,labelindent=0pt,noitemsep,topsep=0pt]
    \item \textbf{Fix Rate} quantifies the number of bugs being fixed by LLMs.
    \item \textbf{Percentage of Plausible Fix} quantifies the number of plausible responses generated by LLMs. The patches of bugs in C\# script files are considered to be \textit{plausible} if they are compilable and pass the bug detection tests by CodeQL; the patches of bugs in asset files are \textit{plausible} if the format is consistent with the YAML language (used for asset files) and passes the test by UnityLint.
    \item \textbf{CodeBLEU}~\cite{ren2020codebleu} is employed to assess the similarity between LLM responses and reference answers from authors. 
We employ hyperparameters demonstrated to have the strongest correlation with human evaluations in existing research, \ie, $\alpha, \beta, \gamma, \delta$ = 0.1, 0.1, 0.4, 0.4, where they are weights to combine standard BLEU, weighted $n$-gram match, syntactic AST match, and semantic data-flow match. 
    \item \textbf{Similarity Scores} are adopted for evaluating fixes on asset files. 
    The average scores of total responses are adopted as their similarity scores.
     
\item
\textbf{Percentage of Correct Patches} quantifies the number of correct patches over the total number of manually inspected bugs. 
\end{itemize}

\hn{Since we aggregate the means of CodeBLEU and similarity scores across all samples, we report the average scores with confidence interval computed with bootstrap sampling with 1,000 resamples; $X_{-Z}^{+Y}$ represents ``95\% of the resampled models yielded a score in the range $[X - Z, X + Y]$''. Furthermore, to make a comparison, we compute whether
the mean difference between scores is statistically significant using the Wilcoxon signed-rank test ($\alpha$=0.05).}
\vspace{-0.3cm}

\hn{
\subsection{Compared Techniques}
\label{compare}
\textbf{Baseline techniques.} So far, no existing approaches are specifically designed for fixing XR performance bugs. Thus, we select three SOTA APR approaches that fix general C\# bugs. Although these methods effectively repair general bugs, they are primarily evaluated on benchmark datasets like Defects4J~\cite{just2014defects4j} (Java) and QuixBugs~\cite{lin2017quixbugs} (Python). We adapt these APR methods to our dataset. For comparison, we ask each APR method to generate five responses per bug scenario and evaluate them using the same metrics described in \S~\ref{metrics}. We keep the $\max\_\operatorname{tokens}$ to be the same as \tool.
\begin{itemize}[leftmargin=*,labelindent=0pt,noitemsep,topsep=0pt]
    \item \textbf{AlphaRepair}~\cite{xia2022less} uses CodeBERT for APR by infilling buggy lines with masked tokens. CodeBERT has demonstrated strong generalization for C\# code generation~\cite{feng2020codebert}. Following its work, we provide the LLM with buggy context for three bug scenarios, masking the buggy lines. Top-5 candidate patches are selected for evaluation.
    \item \textbf{CodeT5} (Fine-tuned for multi-hunk APR Tasks~\cite{huang2025comprehensive}) was fine-tuned on the CPatMiner~\cite{li2022dear} dataset to assess LLM's performance in multi-hunk bug repair. Trained on C/C\# code from GitHub, CodeT5~\cite{wang2021codet5} is capable of generalizing to our task.
    \item \textbf{Self-Repair}~\cite{olausson2023demystifying} uses test failure feedback to prompt LLMs for brief failure explanations, then leverages this information to guide code improvements. Here, we use error messages from our SAT tools to generate feedback. To make a fair comparison, we choose GPT-4o as both its feedback and repair models. 
\end{itemize}
}

\vspace{-0.3cm}
\subsection{Research Questions}
We evaluate the effectiveness of the proposed \tool in repairing XR-related bugs and answer the following research questions (\textbf{RQs}):
\begin{enumerate}[leftmargin=*,labelindent=0pt,noitemsep,topsep=0pt,start=1,label={\bfseries RQ\arabic*:}]
   \setlength{\parskip}{0pt}
   \setlength{\itemsep}{0pt}
    \item What is the distribution of XR bugs in our constructed XR codebase? 
    \item What is the performance of XR-related bug detection based on customized static analyzers?
    \item What is the performance of different LLMs in bug-fixing tasks in XR programs?
    \item How should prompts be engineered to instruct LLMs to fix XR-related bugs?
    \item How do LLMs perform under different bug-fixing scenarios?
    \item How is \tool's repairing ability compared to baseline APR methods?
\end{enumerate}


\vspace{-0.3cm}
\subsection{RQ1: Distribution of XR Program Repair Evaluation Dataset}
\begin{figure}[t]
\vspace{-0.3cm}
\centering
\setlength{\abovecaptionskip}{0.1cm}
\subfigure[Distribution of bugs in XR projects]{
\includegraphics[height=2.47cm]{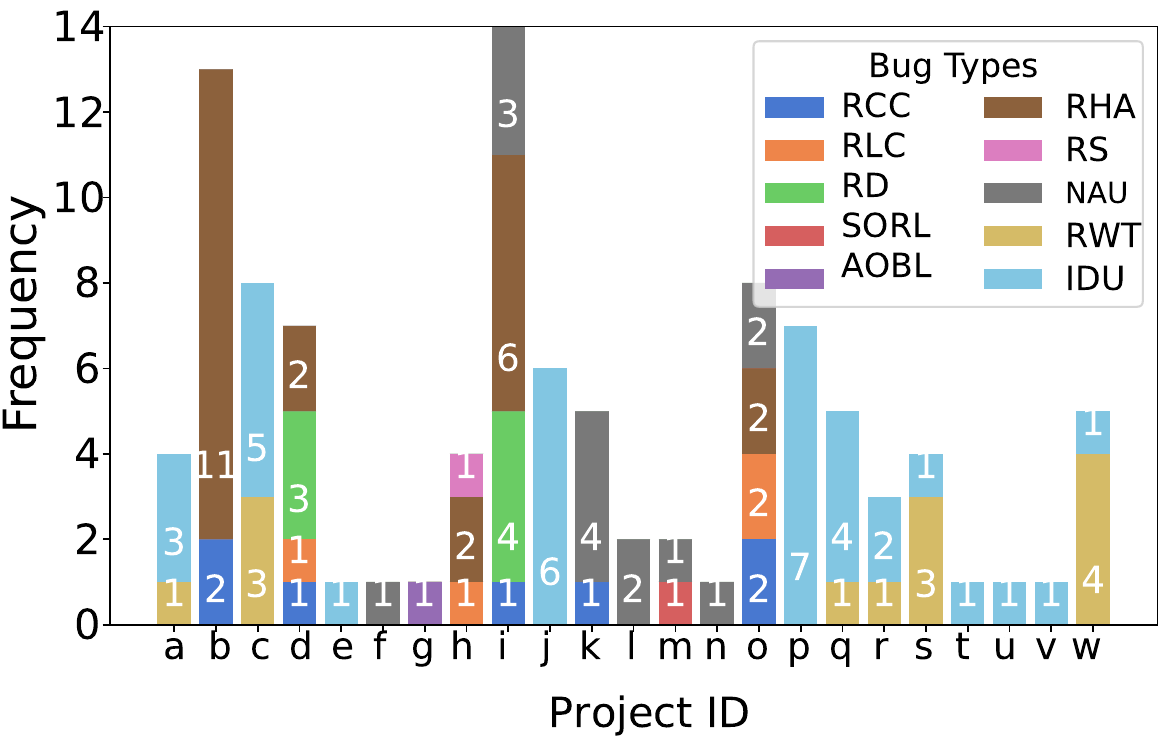}
\label{fig:dataset}
}
\subfigure[Three types of code scenarios]{
\includegraphics[height=2.45cm]{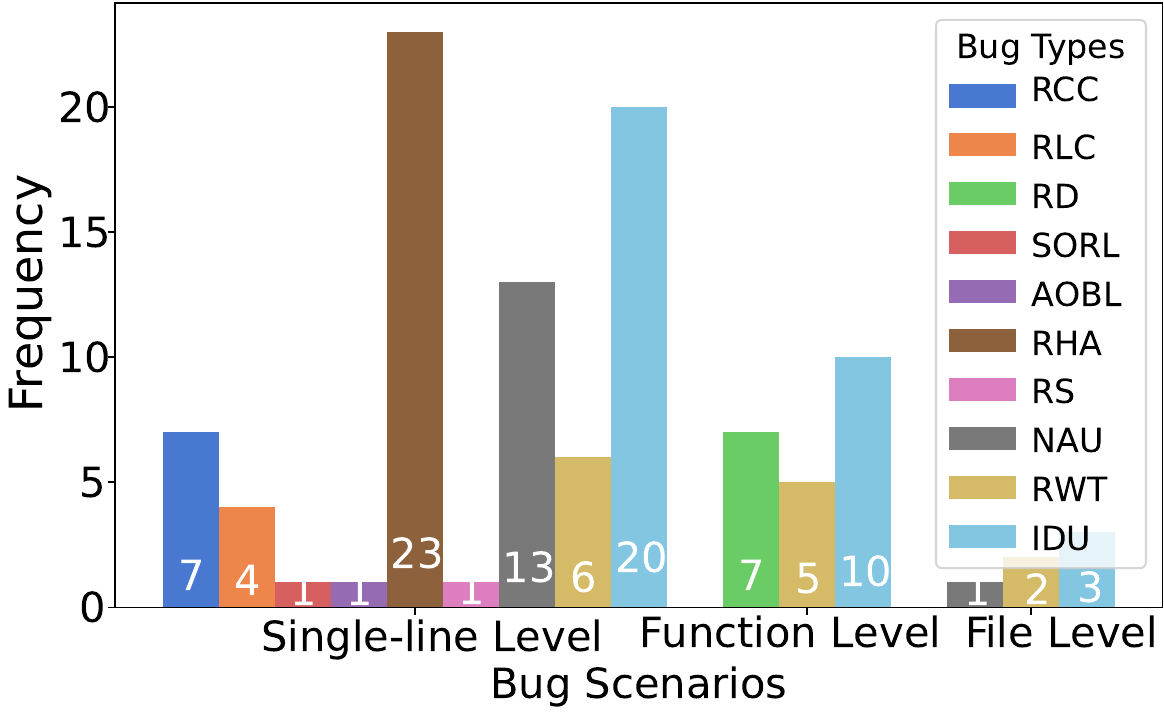}
\label{fig:scenarioinfo}
}
\caption{ Distributions of XR bugs and three code scenarios.}
\label{fig:dis-bug}
\vspace{-0.2cm}
\end{figure}

As shown in \S~\ref{bugdetect}, we curate different bug scenarios based on open-source XR codebases.
Figure~\ref{fig:dis-bug} depicts the distributions of bugs in XR projects and three types of bug scenarios. Our curated dataset contains a total of 104 bugs.
We observe from Figure~\ref{fig:dataset} that the \texttt{\small IDU} bug occupies the largest amount among all bugs, while \texttt{\small SORL}, \texttt{\small AOBL}, and \texttt{\small RS} have the smallest ones. \hn{We also conclude that \texttt{\small IDU} and \texttt{\small RWT} are the most universal bugs among the XR projects, suggesting that these two bugs are pervasive challenges for XR developers. }

Figure~\ref{fig:scenarioinfo} shows the distributions of bugs in three types of bug scenarios, demonstrating the prevalence of different bug scenarios. Among them, single-line level bug scenarios have the highest frequency of 76, \ie, more than 73.07\%. In contrast, function-level and class-level bug scenarios have frequencies of 22\% and 6\%, respectively. \hn{XR's real-time rendering characteristics and frequent interactions make single-line performance bugs visible. }Bugs \texttt{\small IDU} and \texttt{\small RWT} consist of all three code scenarios. Bug \texttt{\small NAU} only contains two code scenarios: single-line level and class-level. The other bugs primarily feature only one scenario. Thus, it is essential for \tool to fix XR performance bugs across diverse code scenarios.

\answer{1}{The curated evaluation dataset provides us with diverse sources of open-source XR code bases with prevalent real-world bugs of different code scenarios.}

\vspace{-0.3cm}
\subsection{RQ2: Performance of Customized Bug Detection Tools}

Bug detection is a prerequisite for bug-fixing tasks. Herein, we evaluate static analysis tools (SATs) on all bugs that we collected to find if there are any false positives. Two authors with professional XR development experience independently analyze all detected bugs to determine precision. Then, they have a two-hour meeting to discuss all detection results. \hn{We employed Cohen's Kappa coefficient to verify how two authors agree on the detection result. Their conclusions achieve strong consistency (0.83>0.8). }Table~\ref{tab:precision} shows the precision of SATs over different rules. For existing rules for detection rules for \texttt{\small RHA}, we have found nine false positives (one for \texttt{\small RCC}). We report them to CodeQL. For other detection rules, we have found no false positives. \hn{Overall, our static analysis tools can achieve a high precision of 90.4\%. }

\begin{table}[t]
\vspace{-0.3cm}
\setlength{\abovecaptionskip}{0.1cm}
\caption{Evaluation of SATs on all detected bugs.}
\label{tab:precision}
\centering
\scalebox{0.7}{
\setlength{\tabcolsep}{1mm}{
    \renewcommand{\arraystretch}{0.9}
\begin{tabular}{lllll}\toprule
\textbf{Approach} & 
\textbf{Rules} & 
\textbf{False Positive} & \textbf{True Positive} & \textbf{Precision}
\\\midrule
\multirow{2}{*}{Our SAT Tools} & Existing & 10 & 34 & \multirow{2}{*}{90.4\%} \\
 & Customized & 0 & 60 & \\
\bottomrule
\end{tabular}}}
\vspace{-0.3cm}
\end{table}
\answer{2}{In summary, the existing rules in UnityLint and CodeQL are reliable in detecting errors with high precision. Moreover, our static analysis tools are effective in detecting XR bugs with a precision of 90.4\%.}

\vspace{-0.3cm}
\begin{figure}[htbp]
\setlength{\abovecaptionskip}{0.1cm}
\centering
\includegraphics[width=8.3cm]{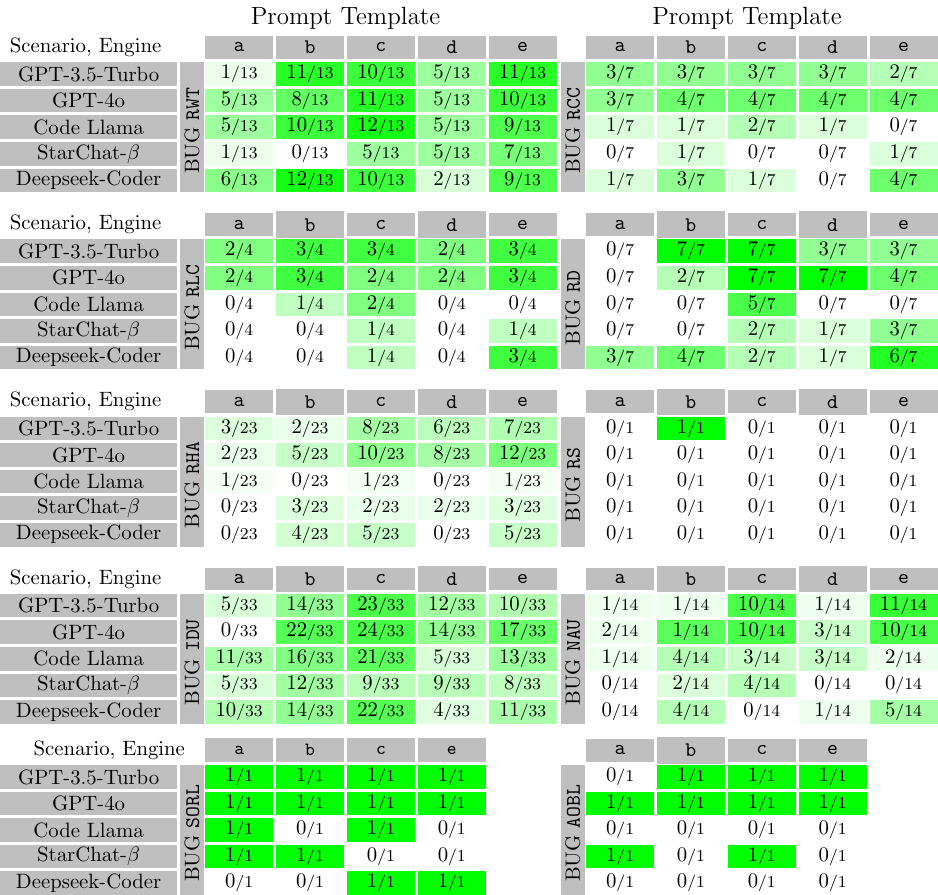}
\caption{\hn{Statistics of fixed bugs by different LLMs instructed by prompt variants.}}
\label{fig:result}
\vspace{-0.6cm}
\end{figure}

\subsection{RQ3: Performance of Different LLMs on XR Program Repair. }\label{subsec:rq3}
\subsubsection{Static Analysis.} We first evaluate the performance of LLMs on bug-fixing tasks via customized static tools. Figure~\ref{fig:result} shows the statistics of different bugs being fixed by diverse LLMs with different prompts. As shown in Figure~\ref{fig:result}, GPT-4o achieves the best performance instructed by prompt c, \ie, generating 70 plausible bug fixes out of 104, with a success fix rate of 67.31\%, slightly higher than GPT-3.5-Turbo's 63.46\%. \hn{For code LLMs, Code Llama achieves its highest fix rate of 45.19\% with prompt c. Deepseek-Coder performs the best with prompt e, reaching a fix rate of 42.31\%.} In contrast, StarChat-$\beta$ exhibits significantly lower bug-fixing capability compared to other Code LLMs, with a fix rate of only 23.08\%. Among all XR bugs, \texttt{\small RHA} is the most challenging one to repair, showing the lowest fix rate of 52.17\%. This challenge arises because \texttt{\small RHA} bugs involve redundant resource allocation, requiring the removal of statements for a fix. LLMs struggle with them because they are less inclined to generate responses that comment out or delete the original buggy code lines.

\begin{figure}[t]
\centering
\includegraphics[width=6cm]{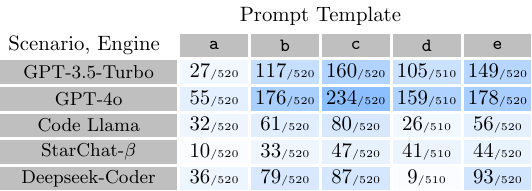}
\caption{{Percentage of Plausible Fix generated by LLMs.}}
\label{fig:compilable}
\vspace{-0.6cm}
\end{figure}

Figure~\ref{fig:compilable} presents the percentage of plausible fixes generated by LLMs. GPT-4o achieves dominating performance on generating plausible fixes instructed by prompt c of 45\%, significantly outperforming GPT-3.5-Turbo's 30.77\%. In contrast, code LLMs have lower plausible fix rates, with Code Llama and Deepseek-Coder achieving 15.38\% and 17.88\%, respectively. Notably, prompt e performs effectively with Deepseek-Coder, producing six more plausible fixes than prompt c. StarChat-$\beta$ performs the worst, with a rate of 9.04\%.

\begin{table*}[h]
\caption{\hn{Bug-fixing performance (Average Score) of LLMs over different bug types in both C\# scripts and asset files.}}
\label{tab:codebleu}
\centering
\scalebox{0.7}{
    \renewcommand{\arraystretch}{0.9}
\begin{tabular}{*{13}{c}} 
\toprule
    \multirow{2}{*}{\textbf{Engine}} &
    \multicolumn{9}{c}{\textbf{CodeBLEU}} & \multicolumn{3}{c}{\textbf{Similarity Scores}} \\ \cline{2-10} 
    \cline{11-13}
      & \cellcolor{mylightblue} \texttt{\small RWT}& \cellcolor{mylightblue} \texttt{\small RCC} & \cellcolor{mylightblue} \texttt{\small RLC} & \cellcolor{mylightblue} \texttt{\small RD} & \cellcolor{mylightblue} \texttt{\small RHA} & \cellcolor{mylightblue} \texttt{\small RS} & \cellcolor{mylightblue} \texttt{\small IDU} & \cellcolor{mylightblue} \texttt{\small NAU} &\cellcolor{mylightblue}  Overall & \cellcolor{mygray} \texttt{\small SORL}& \cellcolor{mygray} \texttt{\small AOBL} & \cellcolor{mygray} Overall\\ \midrule
      {GPT-3.5-Turbo} & \cellcolor{mylightblue} \textbf{0.831} & \cellcolor{mylightblue} \underline{0.610} & \cellcolor{mylightblue} \textbf{0.887}  & \cellcolor{mylightblue} \underline{0.631}  & \cellcolor{mylightblue} \underline{0.558} & \cellcolor{mylightblue} \textbf{0.759}  & \cellcolor{mylightblue} \underline{0.507}
      &\cellcolor{mylightblue}  {0.248}&\cellcolor{mylightblue} \underline{0.560} &\cellcolor{mygray}0.2&\cellcolor{mygray}0.4 &\cellcolor{mygray}\underline{0.3}   \\ 
      {GPT-4o} &\cellcolor{mylightblue} \underline{0.760}  &\cellcolor{mylightblue} \textbf{0.622} &\cellcolor{mylightblue} \underline{0.885} &\cellcolor{mylightblue} \textbf{0.746}  &\cellcolor{mylightblue} \textbf{0.651} &\cellcolor{mylightblue} \underline{0.722} &\cellcolor{mylightblue} \textbf{0.567} &\cellcolor{mylightblue} 0.262 &\cellcolor{mylightblue} \textbf{0.607} &\cellcolor{mygray} \textbf{1.0}  &\cellcolor{mygray} \textbf{1.0} &\cellcolor{mygray}  \textbf{1.0}  \\ 
      {Code Llama} &\cellcolor{mylightblue} 0.636  &\cellcolor{mylightblue} 0.532 &\cellcolor{mylightblue} 0.539 &\cellcolor{mylightblue} 0.620 &\cellcolor{mylightblue} 0.440 &\cellcolor{mylightblue} 0.358 &\cellcolor{mylightblue} 0.472 &\cellcolor{mylightblue} \underline{0.295} &\cellcolor{mylightblue} 0.485 &\cellcolor{mygray} \underline{0.4}  &\cellcolor{mygray} 0 &\cellcolor{mygray} 0.2 \\ 
      {StarChat-$\beta$} &\cellcolor{mylightblue} 0.472 &\cellcolor{mylightblue} 0.462 &\cellcolor{mylightblue} 0.437 &\cellcolor{mylightblue} 0.500 &\cellcolor{mylightblue} 0.429  &\cellcolor{mylightblue} 0.265 &\cellcolor{mylightblue} 0.392 &\cellcolor{mylightblue} \textbf{0.307} &\cellcolor{mylightblue} 0.418 &\cellcolor{mygray} 0 &\cellcolor{mygray} \underline{0.6} &\cellcolor{mygray}  \underline{0.3}   \\ 
      {Deepseek-Coder}   
      &\cellcolor{mylightblue} 0.626  &\cellcolor{mylightblue} 0.492 &\cellcolor{mylightblue} 0.531 &\cellcolor{mylightblue} 0.449  &\cellcolor{mylightblue} 0.441 &\cellcolor{mylightblue} 0.441 &\cellcolor{mylightblue} 0.432 &\cellcolor{mylightblue} 0.282 &\cellcolor{mylightblue} 0.446 &\cellcolor{mygray} \underline{0.4}  &\cellcolor{mygray} 0 &\cellcolor{mygray} 0.2
      \\ 
      
      \bottomrule
\end{tabular}
}
\end{table*}
\vspace{-0.3cm}

\subsubsection{Reference Answer Comparison.}
We evaluate the performance of LLMs through reference answer comparison. \hn{For C\# script files, GPT-4o and GPT-3.5-Turbo achieve  overall CodeBLEU averages of $0.608_{-.0095}^{+.0090}$ and $0.560_{-.0097}^{+.0099}$, respectively, as detailed in Table~\ref{tab:codebleu}, indicating a statistical difference ($P$-value<0.05). Among code LLMs, Code Llama scores $0.485_{-.0077}^{+.0073}$, with Deepseek-Coder and StarChat-$\beta$ achieving $0.445_{-.0074}^{+.0074}$ and $0.418_{-.0068}^{+.0066}$, respectively. There is a statistical difference observed between Code Llama and Deepseek-Coder, both of which significantly outperform StarChat-$\beta$.} For bug fixing in asset files, GPT-4o performs the best on \texttt{\small SORL} and \texttt{\small AOBL}, with an overall average score of 1.0. Some bugs, like \texttt{\small RLC},   \texttt{\small SORL}, and \texttt{\small AOBL}, achieve higher CodeBLEU and similarity scores, exceeding 0.8. Notably, \texttt{\small NAU} is the most challenging bug for LLMs to generate answers similar to the reference answer, with StarChat-$\beta$ achieving the highest score of 0.307. This is because LLMs tend to introduce new variables located different from the reference solutions.


\vspace{-0.2cm}
\subsubsection{Manual Inspection.}
Besides static analysis and reference answer comparison, we also perform manual inspections to check whether XR projects can be successfully executed after bugs are fixed. However, this process is extremely laborious and demanding when considering the huge XR project size. To this end, we select 8 out of 10 bugs, in which each LLM generates one plausible fix, instructed by the same prompt variants to analyze their responses. We manually check the correctness of the generated plausible patches by selecting one case per bug where LLMs produce plausible fixes. For each case, we check at most 10 responses among all plausible patches. During this process, one professional researcher evaluates these responses carefully and discusses them with another researcher to make a decision. Their conclusions also reach strong consistency with Cohen's Kappa coefficient score of 0.9. We employ the percentage of correct responses to represent manual inspection results. Table~\ref{tab:inspect} shows that GPT-4o produces 8.33\% more correct patches than GPT-3.5-Turbo. Meanwhile, Deepseek-Coder performs better than other Code LLMs. A large number of answers generated by Code Llama and StarChat-$\beta$ leave out the original error code lines. Thus, their fixes are unreliable in this sense.

To further evaluate the reliability of LLMs, we select one scenario for \texttt{\small SORL} and \texttt{\small AOBL}, where GPT-4o generates at least one plausible fix. \hn{We choose these two bugs to investigate the performance issues quantitatively and qualitatively.} After replacing the original program with the response generated by LLMs, we compile and execute the entire project in the Unity Editor to test its functionality. For \texttt{\small SORL}, GPT-4o suggests developers change the settings of ``\texttt{\small m\_Lightmapping}'' to be 2, for scenes with only static objects. After adjusting this setting in the Unity Editor and running the scene in Play Mode, the frame rendering time, 
largely decreases from 2.4 ms to 1.8 ms by 33\%, indicating the successful fix of this bug. For \texttt{\small AOBL}, GPT-4o recommends developers adjust the settings to real-time values due to the presence of animation components in the scene. This fix allows the lighting effect in the render in real-time, thereby greatly enhancing the realism and vividness, especially when objects are in motion.

\answer{3}{The evaluation results demonstrate LLMs' great potential in fixing XR bugs. General LLMs perform better than Code LLMs. For general LLMs, {\footnotesize GPT-4o} surpasses {\footnotesize GPT-3.5-Turbo} in terms of correctness and reliability when addressing XR bugs. For Code LLMs, {\footnotesize Code Llama} and {\footnotesize Deepseek-Coder} show better performance compared to {\footnotesize StarChat-$\beta$} while {\footnotesize Deepseek-Coder} generates more correct patches than {\footnotesize Code Llama} according to manual inspection.}

\begin{table}[h]
\setlength{\abovecaptionskip}{0.1cm}
\caption{ Percentage of Correct Patches of Manual Inspection.}
\label{tab:inspect}
\centering
\scalebox{0.65}{
\renewcommand{\arraystretch}{0.8}
\begin{tabular}{*{6}{c}} 
\toprule
    \multirow{2}{*}{\textbf{Bug Type}} & \multicolumn{5}{c}{\textbf{LLM Engines}}
   \\ \cmidrule(l){2-6}
      & \textbf{GPT-3.5-Turbo} & \textbf{GPT-4o} & \textbf{Code Llama} &  \textbf{StarChat-$\beta$} & \textbf{Deepseek-Coder}  \\ \midrule
      \texttt{\small RWT} & 50 & 50 & 33 &  0 & 20\\ 
      \texttt{\small RCC} & 50  & 50 & 0 & 0 & 20\\ 
      \texttt{\small RLC} & 100  & 100 & 100 & 0 & 100\\ 
      \texttt{\small RD} & 100 & 100 & 0 & 0 & 0\\ 
      \texttt{\small RHA} & 100 & 100 & 0 & 100 & 100\\ 
      \texttt{\small IDU} & 100 & 100 & 50 & 0 & 50\\ 
      \texttt{\small NAU} & 0 & 50 & 0 & 0 & 50\\ 
      \texttt{\small SORL} & 100 & 100 & 100 & 100 & 100\\  \midrule
      Average & 75 & \textbf{81.25} & 35.38 & 25 & 55\\
      \bottomrule
\end{tabular}}
\vspace{-0.5cm}
\end{table}

\subsection{RQ4: Evaluation on Prompt Templates in Program Repair.}

\subsubsection{LLMs' Performance with Different Prompt Templates}
As stated before, most of the LLMs achieve their best performance when instructed by prompt c regarding the fix rate and the percentages of plausible responses. We next compare the LLM-generated responses of different prompt templates with their reference answers. 

\begin{figure}[h]
\setlength{\abovecaptionskip}{0.1cm}
	\centering
    \includegraphics[width=.37\textwidth]{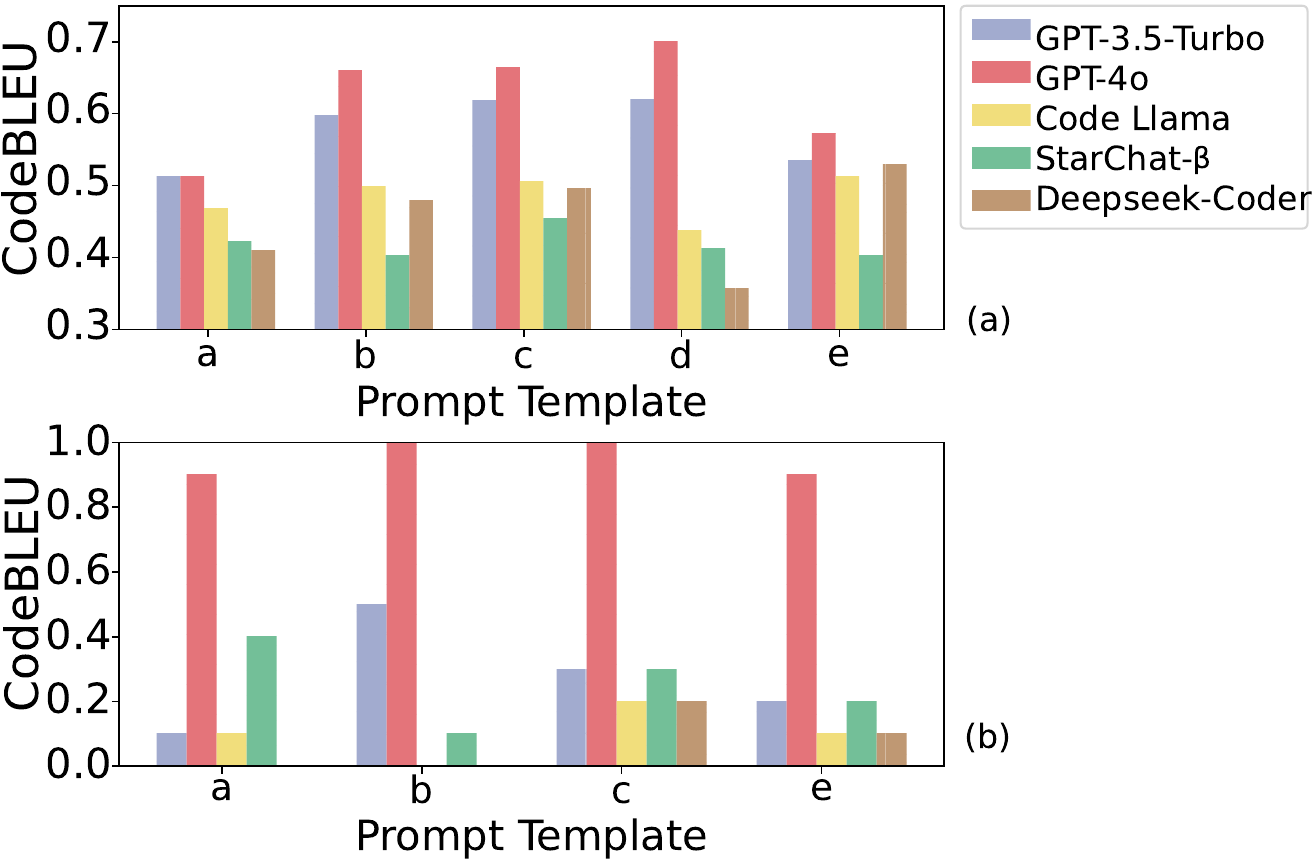}
    \caption{\hn{Reference Answer Comparison of different LLMs instructed by prompt variants.} }
    \label{fig:prompt}
    \vspace{-0.5cm}
\end{figure}

Figure~\ref{fig:prompt} presents the results of LLMs instructed by different prompt templates in script and asset files in terms of the reference answer comparison.
First, Figure~\ref{fig:prompt}(a) plots  LLMs' average CodeBLEU scores in fixing bugs in C\# scripts instructed by five prompts. \hn{For GPT-4o, it shows significantly better performance instructed by prompt d ($0.700_{-.0194}^{+.0207}$) than prompt c ($0.664_{-.0196}^{+.0189}$) with $P$-value < 0.05. For GPT-3.5-Turbo, prompts c and d show no significant difference with
$P$-value of 0.44. We conclude that prompt d is more efficient for GPT-4o to  produce fixes similar to reference answers. For code LLMs, both Code Llama and Deepseek-Coder achieve their highest scores with prompt e. In contrast, StarChat-$\beta$ performs the best with prompt c. }
 Figure~\ref{fig:prompt}(b) reports similarity scores of LLMs in fixing bugs in asset files. Among all LLMs, GPT-4o, Code Llama, and Deepseek-Coder achieve the highest similarity scores instructed by prompt c. Differently, GPT-3.5-Turbo and StarChat-$\beta$ achieve the best performance instructed by prompt b and prompt a. 

\vspace{-0.2cm}

\hn{\subsubsection{Ablation Study.}
\label{abla}
We conduct an ablation study to examine the impacts of incorporating different contextual components in our prompt templates (in \S~\ref{sec:programrepair}). We assess the impacts by calculating the mean scores with standard deviation across five LLMs using three metrics: Fix Rate (FR), Percentage of Plausible Fix (PPF), and CodeBLEU. Table~\ref{tab:ablation} lists the results of different prompt components' contribution, where \textbf{Basic Instruction} serves as the baseline.}

\hn{We observe that the \textbf{Bug Instruction} component contributes the highest improvement across all metrics. \textbf{Fix Instruction} also enhances performance, but to a less degree. The results indicate that \textit{highlighting a bug} is more effective than simply suggesting how to fix it. In contrast, \textbf{Alternative Comment Style} lowers performance, particularly the fix rate. The high CodeBLEU deviation also indicates LLMs are less consistent with this format. The \textbf{Code Examples} component offers modest gains in fix rate and plausible fixes, but may slightly reduce CodeBLEU. This may be because the model relies too much on examples instead of actually addressing the bug.} \hn{In summary, providing clear and relevant contextual information, especially bug descriptions, significantly improves bug-fixing performance. Although code examples help LLMs suggest more possible fixes, they are not always reliable. Changing comment styles can impair effectiveness.}

\begin{table}[t]
\setlength{\abovecaptionskip}{0.1cm}
\caption{\hn{Prompt Component Contribution.} }
\label{tab:ablation}
\centering
\scalebox{0.75}{
\setlength{\tabcolsep}{1mm}{
\renewcommand{\arraystretch}{0.8}
\begin{tabular}
{llp{2.5cm}ll}\toprule
\textbf{Prompt Components} & 
\textbf{FR (\%)} & 
\textbf{PPF (\%)} & \textbf{CodeBLEU} 
\\\midrule
Basic Instruction & 15.4 ($\pm$4.21) & 6.15 ($\pm$2.79) & 0.465 ($\pm$.0436) \\
\midrule
Bug Instruction  & \textbf{+19.6} ($\pm$7.63) & \textbf{+11.8} ($\pm$7.31) & \textbf{+0.0628} ($\pm$.0558) \\
Fix Instruction & +12.9 ($\pm$8.72) & +5.46 ($\pm$3.65) & +0.0200 ($\pm$.0170) \\
Alternative Comment Style & -24.9 ($\pm$10.1) & -10.1 ($\pm$4.91) & -0.0424 ($\pm$.0601) \\
Code Example & +4.04 ($\pm$6.76) & +2.08 ($\pm$2.41) & -0.0174 ($\pm$.0504) \\
\bottomrule
\end{tabular}}}
\vspace{-0.3cm}
\end{table}

\answer{4}{To conclude, prompt c helps most  LLMs generate more plausible fixes. The ablation study indicates that adding relevant instructions especially bug instruction improves performance. }

 \vspace{-0.3cm}
\subsection{RQ5: Performance of LLMs in different bug scenarios.}
We next evaluate the performance of LLMs in different bug scenarios. Table~\ref{tab:bugscenario} reports the results in three types of bug scenarios instructed by prompt c. For single-line level bugs, GPT-4o generates the highest number of plausible patches, with 199 out of 380. For function-level bugs, GPT-4o leads with 25 out of 110 fixes. It also ranks first for class-level bugs, producing 10 out of 30 fixes. While GPT-3.5-Turbo ranks second for single-line and function-level bugs, Code Llama takes the second spot for class-level bugs with 6 out of 30. To conclude, we demonstrate that GPT-4o is more suitable to be integrated into our \tool framework compared to other LLMs.

\answer{5}{LLMs perform differently when handling different bug scenarios. {\footnotesize GPT-4o} outperforms other LLMs in fixing all bug scenarios.}

\begin{table}[t]
\vspace{-0.1cm}
\setlength{\abovecaptionskip}{0.2cm}
\caption{Plausible responses generated by LLMs in different bug scenarios using prompt c.}
\label{tab:bugscenario}
\centering
\scalebox{0.7}{
\renewcommand{\arraystretch}{0.8}
\begin{tabular}{*{4}{c}} 
\toprule
     \multirow{2}{*}{\textbf{Engine}} &
    \multicolumn{3}{c}{\textbf{Bug Scenarios}} \\ \cmidrule(l){2-4}
       & Single-line level & Function level & Class level \\ \midrule
      {GPT-3.5-Turbo} & \underline{132} &\underline{24} & 4   \\ 
      {GPT-4o} & \textbf{199}  & \textbf{25} & \textbf{10} \\ 
      {Code Llama} & 59  & 15 &  \underline{6} \\ 
      {StarChat-$\beta$} & 40  & 5 & 2   \\ 
      {Deepseek-Coder} & 68&16 & 3\\
\bottomrule
\end{tabular}}
\vspace{-0.4cm}
\end{table}

\begin{table}[t]
\setlength{\abovecaptionskip}{0.1cm}
\caption{\hn{Bug Repair Performance of \tool vs SOTA APR Methods.  ($\ddagger$ indicates no statistically significant difference.) }}
\label{tab:compare}
\centering
\scalebox{0.75}{
\setlength{\tabcolsep}{1mm}{
\renewcommand{\arraystretch}{0.8}
\begin{tabular}
{llp{2.5cm}l}\toprule
\textbf{Approach} & 
\textbf{FR (\%)} & 
\textbf{PPF (\%)} & \textbf{CodeBLEU} 
\\\midrule
AlphaRepair & 5.77 & 2.74 & 0.404\\
Fine-tuned CodeT5 & 6.73 & 3.37 & 0.375\\
Self-Repair (GPT-4o) & 59.6  & 41.5 & \textbf{0.667}$\ddagger$ \\ \midrule
\textbf{\tool} (GPT-4o) & \textbf{67.3} & \textbf{45.0} & 0.664$\ddagger$ \\ 
\bottomrule
\end{tabular}}}
\end{table}
\vspace{-0.5cm}

\hn{\subsection{RQ6: \tool's Repair Ability Compared with SOTA APR Methods.}
Table~\ref{tab:compare} compares our \tool with SOTA APR methods. We use the best-performing LLM (GPT-4o) instructed by prompt c. Table~\ref{tab:compare} shows that \tool significantly outperforms the baselines. While AlphaRepair and Fine-tuned CodeT5 both fix fewer than 7\% of bugs, \tool achieves a 67.3\% fix rate and surpasses the runner-up approach by 7.7\%. \tool also attains the highest percentage of plausible fixes, and its CodeBLEU of 0.664 demonstrates that higher correctness is achieved without decreasing code similarity. Specifically, AlphaRepair and fine-tuned CodeT5 struggle to handle diverse XR code scenarios. Meanwhile, Self-Repair fails because current LLMs cannot reliably or accurately explain coding errors.~\cite{zhang2024systematic}.}

\answer{6}{\tool's repair ability has proven to be more effective in producing correct and reliable fixes than all three APR methods.}

\vspace{-0.3cm}
\subsection{Case Study and Discussion}
\subsubsection{Case Study.}
\begin{figure}[h]
\vspace{-0.3cm}
\centering
\includegraphics[width=1\linewidth]{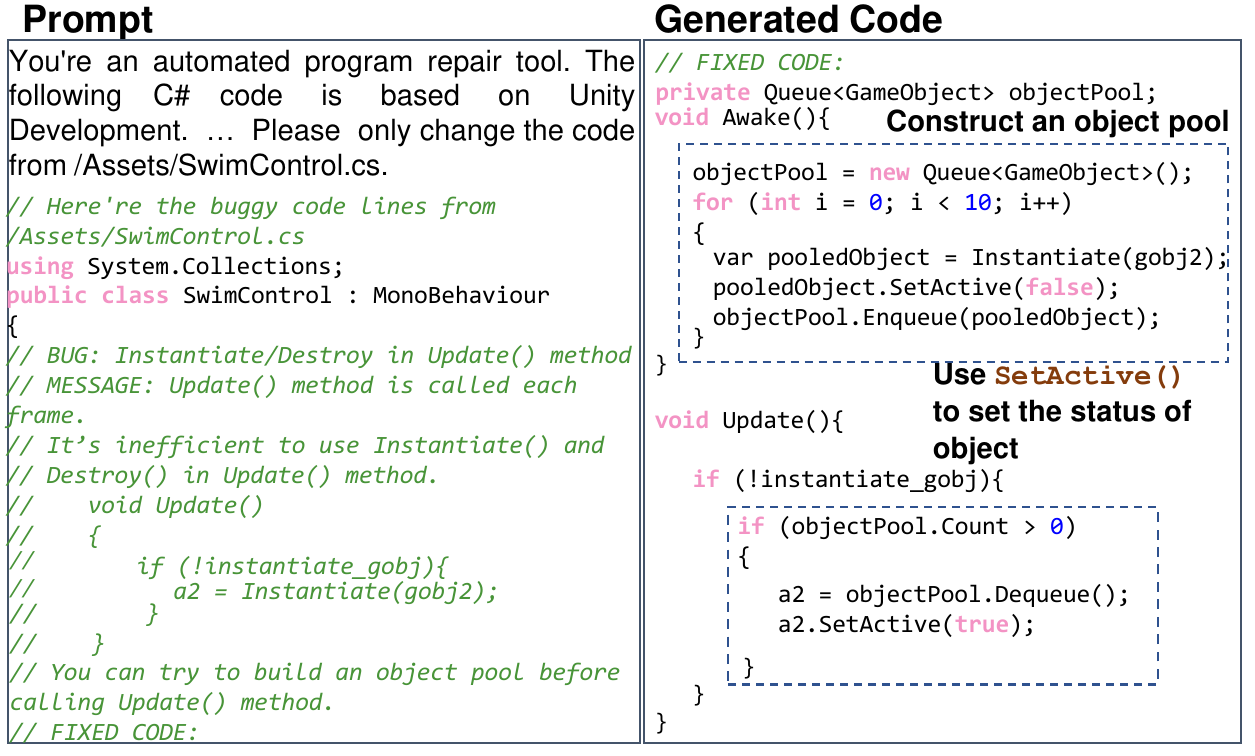}
\caption{
One Example of Fixing \texttt{\small IDU} Bug using GPT-4o.}
\label{fig:casestudy}
\centering
\vspace{-0.3cm}

\end{figure}

We provide a case study of using GPT-4o instructed by prompt c to fix \texttt{\small IDU}. Figure~\ref{fig:casestudy} illustrates the comprehensive procedure for instructing LLMs to fix the bug according to the bug detection information provided by CodeQL. In the left column, we provide GPT-4o with buggy code lines accurately located by the designed queries of CodeQL. Then, we comment out the contents using ``//'' and add a description of how to fix this bug. The right column exhibits the generated code by GPT-4o, which first constructs a pool with 10 objects following the fix instructions. Next, it utilizes \texttt{\small SetActive()} to control the GameObject's visibility. In this way, it avoids creating objects frequently in \texttt{\small Update()} function, consequently repairing the bug. 

\vspace{-0.3cm}
\hn{\subsubsection{Discussion.} 
For practical applications, we highlight several key considerations. (1) \textit{Effectiveness}: General LLMs generally demonstrate superior performance to Code LLMs, with GPT-4o surpassing all other LLMs in effectiveness. (2) \textit{Bug Type Specificity}: LLMs show different abilities with various bug types. Notably, GPT-3.5-Turbo is not effective for class-level bugs because it cannot handle them well. (3) \textit{Code LLM Performance}: While Code Llama achieves the highest fix rate, Deepseek-Coder consistently yields more dependable results among all code LLMs. (4) \textit{Prompt Design}: Prompt c tends to generate more plausible code fixes, whereas prompt d produces more reliable outputs.}

\textbf{Threats to Internal Validity.} 
The performance of the proposed \tool depends on customized static analysis tools, such as UnityLint and CodeQL. To ensure the reliability of static analysis in bug detection, we examined the correctness of pre-defined and customized rules in UnityLint and CodeQL. We find that the precision is high. With respect to another internal validity threat caused by the effectiveness of fixing bugs by LLMs, we also conduct static analysis, reference answer comparison, and manual inspection to ensure that fixes generated by LLMs really work in XR projects. 

\textbf{Threats to External Validity.} Our \tool mainly covers ten typical types of XR bugs. Moreover, our study mainly focuses on XR apps built on Unity while there exist many other development frameworks, such as Unreal Engine~\cite{qiu2016unrealcv}, SteamVR~\cite{steamvr}, \etc. These threats may limit the scalability of our approach. 

Moreover, a considerable number of our collected XR bugs are contained in these projects committed before October 2023, i.e., the latest cut-off date among all chosen LLMs. As a result, it is likely that our experimental results may slightly suffer from data leakage. To mitigate this issue, we select all bugs from our curated dataset that were committed after the cut-off date and end up gathering nine bugs. After examining the results, we found that all nine bugs can still be fixed by LLMs. Further, the trends observed in these bugs were consistent with the entire dataset. This result provides evidence that our conclusions still hold even with low data leakage.

 \textbf{Threats to Construct Validity.} \tool lacks bug-fixing capabilities for visual and non-visual functions in XR apps, such as stereoscopic visual inconsistency bugs as reported in~\cite{10.1145/3660803}. To address this threat, we will further extend \tool with other tools (such as large multi-modality models) to support these features. 

\vspace{-0.3cm}
\section{Related Work}
\textbf{XR development bugs.} Most existing studies mainly focus on the development of XR apps while overlooking bad practices or bugs. As one of the earliest studies, Adams \etal~\cite{adams2018ethics} raised concerns about the potential risks of VR apps by interviewing 20 VR users and developers. Li \etal~\cite{li2020exploratory} categorized WebXR bugs based on real-world bug scenarios collected from GitHub and also compared the collected bugs with other conventional Web XR apps. Recently, Rzig et al., ~\cite{rzig2023virtual} identified the phenomenon of lacking automatic tests on open-source VR apps. 
Nusrat \etal~\cite{9402052} manually categorized the optimizations of XR apps into 11 different types and applied static analysis tools to investigate their influence on different life-cycle stages of VR apps. 
Bosco \etal~\cite{10174134} proposed \textit{UnityLint} to detect different types of bad smells for Unity game development by extending Borrelli \etal~\cite{borrelli2020detecting}'s previous framework.

\textbf{Automated program repair.}
Recently, research endeavors have been made to APR \cite{gazzola2018automatic, monperrus2018automatic}. 
In the early stage, search-based APR generates potential fixes and then utilizes heuristic methods~\cite{le2011genprog} and generic programming~\cite{jobstmann2005program} to identify the correct patches. 
Following these attempts, constraint-based methods were introduced to use specifications that guide the repair process, such as Nopol~\cite{xuan2016nopol} and SemFix~\cite{nguyen2013semfix}. 
Moreover, template-based APR approaches were proposed to design specific fix templates for generating fix patches~\cite{le2016history, kim2013automatic, koyuncu2020fixminer, koyuncu2019ifixr}. \hn{Deep learning has greatly improved APR by allowing the use of empirical knowledge.
Tufano et al.~\cite{tufano2019empirical} first established bug-fix pairs (BFPs) and built an NMT model to learn related knowledge. 
APR methods have evolved through the integration of LLMs. For instance, AlphaRepair~\cite{xia2022less} and TypeFix~\cite{peng2024domain} treat the repair task as the infilling task, constructing different prompt templates to repair. Besides these methods, recent studies have explored the effectiveness of zero-shot and few-shot learning for APR using LLMs~\cite{fan2023automated, Hammond2023}. Additionally, Olausson et al.~\cite{olausson2023demystifying} enhance LLM-based repair by enabling self-repair, where models iteratively improve LLM-generated code with feedback.
For fine-tuning, Mashhadi et al. \cite{mashhadi2021applying} first used fine-tuned CodeBERT to repair single-line bugs. In our work, we chose fine-tuned CodeT5~\cite{huang2025comprehensive} to repair multi-hunk bugs as a baseline. More recently, agent-based APR approaches have gained popularity. FixAgent~\cite{lee2024unified}, RepairAgent~\cite{bouzenia2025repairagent}, and Agentless~\cite{xia2025demystifying} allow LLMs to assume multiple roles and interact with external tools to facilitate the repair process. }

\textbf{Pre-trained Language Models for Code.}
Pretrained language models (LMs) with millions to billions of parameters have been proposed and adapted to perform code-related tasks. Most of them can be classified into three groups: encoder-only, decode-only, and encoder-decoder LLMs. Encoder-only LLMs, such as CodeBERT~\cite{feng2020codebert} and GraphCodeBERT~\cite{guographcodebert} are mainly pre-trained using Masked Language Modeling (MLM) tasks on a large corpus. They are mainly applied in code understanding tasks such as code search. Encoder-Decoder LLMs, such as CodeT5~\cite{wang2021codet5}, PLBART~\cite{ahmad2021unified}, and T5~\cite{raffel2020exploring} are suitable for program repair tasks with their sequence-to-sequence settings. Decoder-only LLMs, including GPT-series models, such as GPT-3.5~\cite{gpt3.5}, ChatGPT~\cite{chatgpt}, GPT-4~\cite{achiam2023gpt} and GPT-4o~\cite{gpt4o}, and open-sourced LLMs, like GPT-NeoX~\cite{blackgpt}, InCoder~\cite{friedincoder}, Code Llama~\cite{roziere2023code}, StarChat~\cite{li2023starcoder} and Deepseek~\cite{bi2024deepseek}, are used to perform program repair with examples or instructions to generate code fixes without fine-tuning procedure~\cite{xia2022less, fan2023automated, Hammond2023, ahmad2023fixing, liu2024llm, bouzenia2025repairagent, xia2024automated}. 

\vspace{-0.3cm}

\section{Conclusion}
We present \tool, an LLM-based framework, to fix performance bugs in open-source XR programs. 
First, we construct program repair evaluation datasets with XR-related bugs and open-source XR projects. To accurately detect and localize these XR bugs in XR programs, we design bug detection rules in static analysis tools. Then, we construct meticulously-designed prompt templates to instruct LLMs to fix bugs. Our investigation involves a comparative study of five state-of-the-art LLMs. \hn{We also compare our tools with three SOTA APR approaches.} Our experimental results show that LLMs have great potential in repairing XR bugs even with zero-shot learning. In future work, we will enhance \tool's code understanding and bug localization in XR projects by exploring LLMs' multimodality and agentic intelligence.



\begin{acks}
    The work is supported by The Seed Funding for Collaborative Research Grants of HKBU (with Grant No. RC-SFCRG/23-24/R2/SCI/06) and SD/COMP
Joint Research Scheme (ID: P0042739).
\end{acks}

\newpage
\bibliographystyle{ACM-Reference-Format}
\bibliography{reference}

\end{document}